\PassOptionsToPackage{table, dvipsnames}{xcolor}
\documentclass[sigconf, nonacm]{acmart}
\pdfoutput=1

\newcommand\vldbdoi{XX.XX/XXX.XX}

\newcommand\vldbavailabilityurl{URL_TO_YOUR_ARTIFACTS}
\newcommand\vldbpagestyle{plain}

\usepackage{graphicx}
\usepackage{tikz}
\usepackage{tikzpeople}
\usepackage[algo2e, linesnumbered, ruled, vlined]{algorithm2e}
\usepackage{caption}
\usepackage[inline]{enumitem}
\usepackage{subcaption}

\usepackage{centernot}
\usepackage{adjustbox}
\usepackage{float}
\usepackage{pgfplotstable}
\pgfplotstableset{
    /color cells/min/.initial=0,
    /color cells/max/.initial=1000,
    /color cells/textcolor/.initial=,
    color cells/.code={%
        \pgfqkeys{/color cells}{#1}%
        \pgfkeysalso{%
            postproc cell content/.code={%
                \begingroup
                \pgfkeysgetvalue{/pgfplots/table/@preprocessed
                  cell content}\value
                \ifx\value\empty
                    \endgroup
                \else
                \pgfmathfloatparsenumber{\value}%
                \pgfmathfloattofixed{\pgfmathresult}%
                \let\value=\pgfmathresult
                \pgfplotscolormapaccess
                    [\pgfkeysvalueof{/color cells/min}:\pgfkeysvalueof{/color
                      cells/max}]
                    {\value}
                    {\pgfkeysvalueof{/pgfplots/colormap name}}%
                \pgfkeysgetvalue{/pgfplots/table/@cell content}\typesetvalue
                \pgfkeysgetvalue{/color cells/textcolor}\textcolorvalue
                \toks0=\expandafter{\typesetvalue}%
                \xdef\temp{%
                    \noexpand\pgfkeysalso{%
                        @cell content={%
                            \noexpand\cellcolor[rgb]{\pgfmathresult}%
                            \noexpand\definecolor{mapped
                              color}{rgb}{\pgfmathresult}%
                            \ifx\textcolorvalue\empty
                            \else
                                \noexpand\color{\textcolorvalue}%
                            \fi
                            \the\toks0 %
                        }%
                    }%
                }%
                \endgroup
                \temp
                \fi
            }%
        }%
    },
    format index/.style={
        columns/index/.style={
            column name={},
            string type,
            postproc cell content/.code={},
            postproc cell  content/.append style={
            @cell content/.add={\bfseries \ttfamily \small}{}
            },
        },
    },
    format userId/.style={
        columns/userId/.style={
            string type,
            postproc cell content/.code={}
        },
    },
    format movieId/.style={
        columns/movieId/.style={
            string type,
            postproc cell content/.code={}
        },
    },
    format tag/.style={
        columns/tag/.style={
            string type,
            postproc cell content/.code={}
        },
    },
    format manufacturer/.style={
        columns/manufacturer/.style={
            string type,
            postproc cell content/.code={}
        },
    },
    format Lineage/.style={
        columns/Lineage/.style={
            string type,
            postproc cell content/.code={}
        },
    },
    format method/.style={
        columns/{Provenance Method}/.style={
            string type,
            postproc cell content/.code={},
            postproc cell  content/.append style={
            @cell content/.add={\bfseries}{}
            },
        },
    },
    format table/.style={
        columns/{Lin. size}/.style={
            postproc cell content/.code={}
        },
        every head row/.style={before row=\toprule,after row=\midrule},
        every last row/.style={after row=\bottomrule},
        every odd row/.style={before row={\rowcolor{gray!10}}},
    },
    format stats-table1-exp1/.style={
        columns/Total/.style={
            postproc cell content/.code={}
        },
        columns/Length-L1/.style={
            column name={L[1]},
            postproc cell content/.code={}
        },
        columns/Length-L2/.style={
            column name={L[2]},
            postproc cell content/.code={}
        },
        columns/Length-L3/.style={
            column name={L[3]},
            postproc cell content/.code={}
        },
        columns/Recall-L1/.style={
            column name={L[1]},
        },
        columns/Recall-L3/.style={
            column name={L[3]},
        },
        every head row/.style={before row=\toprule & & \multicolumn{1}{c}{\textbf{Target Tuple}} & \multicolumn{4}{c}{\textbf{Lineage Size}} & \multicolumn{4}{c}{\textbf{Precision}} & \multicolumn{2}{c}{\textbf{Recall}}\\
        \cmidrule(lr){3-3} \cmidrule(lr){4-7} \cmidrule(lr){8-11} \cmidrule(lr){12-13}\\, after row=\midrule},
        every last row/.style={after row=\bottomrule},
        every odd row/.style={before row={\rowcolor{gray!10}}},
    },
    format stats-table2-exp1/.style={
        columns/Total/.style={
            postproc cell content/.code={}
        },
        columns/Length-L1/.style={
            column name={L[1]},
            postproc cell content/.code={}
        },
        columns/Length-L2/.style={
            column name={L[2]},
            postproc cell content/.code={}
        },
        columns/Length-L3/.style={
            column name={L[3]},
            postproc cell content/.code={}
        },
        columns/Recall-L1/.style={
            column name={L[1]},
        },
        columns/Recall-L2/.style={
            column name={L[2]},
        },
        every head row/.style={before row=\toprule & & \multicolumn{1}{c}{\textbf{Target Tuple}} & \multicolumn{4}{c}{\textbf{Lineage Size}} & \multicolumn{4}{c}{\textbf{Precision}} & \multicolumn{2}{c}{\textbf{Recall}}\\
        \cmidrule(lr){3-3} \cmidrule(lr){4-7} \cmidrule(lr){8-11} \cmidrule(lr){12-13}\\, after row=\midrule},
        every last row/.style={after row=\bottomrule},
        every odd row/.style={before row={\rowcolor{gray!10}}},
    },
    format analyst/.style={
        columns/tuple-id/.style={
            column name={Tuple ID},
            string type,
            postproc cell content/.code={}
        },
        columns/related-table/.style={
            column name={Origin Table},
            string type,
            postproc cell content/.code={},
            postproc cell  content/.append style={
            @cell content/.add={\ttfamily}{}
            },
        },
        columns/lineage-level/.style={
            column name={Lineage Level(s)},
            string type,
            postproc cell content/.code={}
        },
        every head row/.style={before row=\toprule & \multicolumn{3}{c}{\textbf{An Analyst's View}} & \multicolumn{2}{c}{\textbf{Hindsight}}\\
        \cmidrule(lr){2-4} \cmidrule(lr){5-6}\\, after row=\midrule},
        every last row/.style={after row=\bottomrule},
        every odd row/.style={before row={\rowcolor{gray!10}}},
    },
    format samples-table/.style={
        every head row/.style={before row=\toprule,after row=\midrule},
        every last row/.style={after row=\bottomrule},
        every odd row/.style={before row={\rowcolor{gray!10}}},
    },
    format samples-nutrients/.style={
        columns/ndb-no/.style={
            column name={\texttt{ndb\_no}},
            string type,
            postproc cell content/.code={}
        },
        columns/derivation-code/.style={
            column name={\texttt{derivation\_code}},
            string type,
            postproc cell content/.code={}
        },
        columns/nutrient-code/.style={
            column name={\texttt{nutrient\_code}},
            string type,
            postproc cell content/.code={}
        },
        columns/nutrient-name/.style={
            column name={\texttt{nutrient\_name}},
            string type,
            postproc cell content/.code={}
        },
        columns/out-val/.style={
            column name={\texttt{out\_val}},
            string type,
            postproc cell content/.code={}
        },
        columns/out-uom/.style={
            column name={\texttt{out\_uom}},
            string type,
            postproc cell content/.code={}
        },
    },
    format stats-table1-exp2/.style={
        columns/name/.style={
            string type,
            postproc cell content/.code={}
        },
        columns/Total/.style={
            postproc cell content/.code={}
        },
        columns/Length-L1/.style={
            column name={L[1]},
            postproc cell content/.code={}
        },
        columns/Length-L2/.style={
            column name={L[2]},
            postproc cell content/.code={}
        },
        columns/Length-L3/.style={
            column name={L[3]},
            postproc cell content/.code={}
        },
        columns/Length-L4/.style={
            column name={L[4]},
            postproc cell content/.code={}
        },
        columns/Recall-L1/.style={
            column name={L[1]},
        },
        columns/Recall-L3/.style={
            column name={L[3]},
        },
        columns/Recall-L4/.style={
            column name={L[4]},
        },
        every head row/.style={before row=\toprule & & \multicolumn{2}{c}{\textbf{Target Tuple}} & \multicolumn{5}{c}{\textbf{Lineage Size}} & \multicolumn{4}{c}{\textbf{Precision}} & \multicolumn{3}{c}{\textbf{Recall}}\\
        \cmidrule(lr){3-4} \cmidrule(lr){5-9} \cmidrule(lr){10-13} \cmidrule(lr){14-16}\\, after row=\midrule},
        every last row/.style={after row=\bottomrule},
        every odd row/.style={before row={\rowcolor{gray!10}}},
    },
    format stats-table2-exp2/.style={
        columns/name/.style={
            string type,
            postproc cell content/.code={}
        },
        columns/Total/.style={
            postproc cell content/.code={}
        },
        columns/Length-L1/.style={
            column name={L[1]},
            postproc cell content/.code={}
        },
        columns/Length-L2/.style={
            column name={L[2]},
            postproc cell content/.code={}
        },
        columns/Length-L3/.style={
            column name={L[3]},
            postproc cell content/.code={}
        },
        columns/Length-L4/.style={
            column name={L[4]},
            postproc cell content/.code={}
        },
        columns/Recall-L1/.style={
            column name={L[1]},
        },
        columns/Recall-L2/.style={
            column name={L[2]},
        },
        columns/Recall-L3/.style={
            column name={L[3]},
        },
        every head row/.style={before row=\toprule & & \multicolumn{2}{c}{\textbf{Target Tuple}} & \multicolumn{5}{c}{\textbf{Lineage Size}} & \multicolumn{4}{c}{\textbf{Precision}} & \multicolumn{3}{c}{\textbf{Recall}}\\
        \cmidrule(lr){3-4} \cmidrule(lr){5-9} \cmidrule(lr){10-13} \cmidrule(lr){14-16}\\, after row=\midrule},
        every last row/.style={after row=\bottomrule},
        every odd row/.style={before row={\rowcolor{gray!10}}},
    },
    format stats-table1-exp3/.style={
        columns/name/.style={
            string type,
            postproc cell content/.code={}
        },
        columns/Total/.style={
            postproc cell content/.code={}
        },
        columns/Length-L1/.style={
            column name={L[1]},
            postproc cell content/.code={}
        },
        columns/Length-L2/.style={
            column name={L[2]},
            postproc cell content/.code={}
        },
        columns/Length-L3/.style={
            column name={L[3]},
            postproc cell content/.code={}
        },
        columns/Recall-L1/.style={
            column name={L[1]},
        },
        columns/Recall-L2/.style={
            column name={L[2]},
        },
        columns/Recall-L3/.style={
            column name={L[3]},
        },
        every head row/.style={before row=\toprule & & \multicolumn{2}{c}{\textbf{Target Tuple}} & \multicolumn{4}{c}{\textbf{Lineage Size}} & \multicolumn{4}{c}{\textbf{Precision}} & \multicolumn{3}{c}{\textbf{Recall}}\\
        \cmidrule(lr){3-4} \cmidrule(lr){5-8} \cmidrule(lr){9-12} \cmidrule(lr){13-15}\\, after row=\midrule},
        every last row/.style={after row=\bottomrule},
        every odd row/.style={before row={\rowcolor{gray!10}}},
    },
    format stats-table2-exp3/.style={
        columns/name/.style={
            string type,
            postproc cell content/.code={}
        },
        columns/Total/.style={
            postproc cell content/.code={}
        },
        columns/Length-L1/.style={
            column name={L[1]},
            postproc cell content/.code={}
        },
        columns/Length-L2/.style={
            column name={L[2]},
            postproc cell content/.code={}
        },
        columns/Length-L3/.style={
            column name={L[3]},
            postproc cell content/.code={}
        },
        columns/Recall-L1/.style={
            column name={L[1]},
        },
        columns/Recall-L2/.style={
            column name={L[2]},
        },
        every head row/.style={before row=\toprule & & \multicolumn{2}{c}{\textbf{Target Tuple}} & \multicolumn{4}{c}{\textbf{Lineage Size}} & \multicolumn{4}{c}{\textbf{Precision}} & \multicolumn{2}{c}{\textbf{Recall}}\\
        \cmidrule(lr){3-4} \cmidrule(lr){5-8} \cmidrule(lr){9-12} \cmidrule(lr){13-14}\\, after row=\midrule},
        every last row/.style={after row=\bottomrule},
        every odd row/.style={before row={\rowcolor{gray!10}}},
    }
}

\theoremstyle{definition}
\newtheorem{runexample}{Running Example}
\newtheorem{example-withrun}{Example}[runexample]

\theoremstyle{definition}
\newtheorem*{runexperiment}{Running Experiments}
\newtheorem{experiment-withrun}{Experiment}

\newcommand{\fncolor}[1]{\textcolor{YellowOrange}{#1}}
\newcommand{\paramcolor}[1]{\textcolor{JungleGreen}{#1}}
\newcommand{\cmtcolor}[1]{\textcolor{Gray}{#1}}

\makeatletter
\newcommand{\removelatexerror}{\let\@latex@error\@gobble}
\makeatother

\usetikzlibrary{calc,trees,positioning,arrows,chains,shapes.geometric,%
    decorations.pathreplacing,decorations.pathmorphing,shapes,%
    matrix,shapes.symbols}

\tikzset{
>=stealth',
  punktchain/.style={
    rectangle, 
    rounded corners, 
    draw=black, very thick,
    text width=22em, 
    minimum height=3em, 
    text centered, 
    on chain},
  line/.style={draw, thick, <-},
  element/.style={
    tape,
    top color=white,
    bottom color=blue!50!black!60!,
    minimum width=8em,
    draw=blue!40!black!90, very thick,
    text width=10em, 
    minimum height=3.5em, 
    text centered, 
    on chain},
  every join/.style={->, thick,shorten >=1pt},
  decoration={brace},
  tuborg/.style={decorate},
  tubnode/.style={midway, right=2pt},
}

\begin{document}
\title{ML Based Lineage in Databases}

\author{Michael Leybovich}
\affiliation{%
  \institution{Technion}
  \city{Haifa}
  \state{Israel}
}
\email{smike87@cs.technion.ac.il}

\author{Oded Shmueli}
\affiliation{%
  \institution{Technion}
  \city{Haifa}
  \state{Israel}
}
\email{oshmu@cs.technion.ac.il}

\begin{abstract}
We track the lineage of tuples throughout their database lifetime. That is, we consider a scenario in which tuples (records) that are produced by a query may affect other tuple insertions into the DB, as part of a normal workflow. As time goes on, exact provenance explanations for such tuples become deeply nested, increasingly consuming  space, and resulting in decreased clarity and readability.
\par We present a novel approach for approximating lineage tracking, using a Machine Learning (ML) and Natural Language Processing (NLP) technique; namely, word embedding. 
The basic idea is summarizing (and approximating) the lineage of each tuple via a small set of constant-size vectors (the number of vectors per-tuple is a hyperparameter). For explicitly (and independently of DB contents)  inserted tuples -  the vectors are obtained via a pre-trained word vectors model over their underlying database domain ``text''. During the execution of a query, we construct the lineage vectors of the final (and intermediate) result tuples in a similar fashion to that of semiring-based exact provenance calculations. We extend the $+$ and $\cdot$ operations to generate sets of lineage vectors, while retaining the ability to propagate information and preserve the compact representation. Therefore, our solution does not suffer from space complexity blow-up over time, and it ``naturally ranks'' explanations to the existence of a tuple.
\par We devise a genetics-inspired improvement to our basic method. The data columns of an entity (and potentially other columns) are a tuple’s basic properties, i.e., the ``genes'' that combine to form its genetic code. We design an alternative lineage tracking mechanism, that of keeping track of and querying lineage (via embeddings) at the column (``gene'') level; thereby, we manage to better distinguish between the provenance features and the textual characteristics of a tuple. Finding the lineage of a tuple in the DB is analogous to finding its predecessors via DNA examination.
\par We further introduce several improvements and extensions to the basic method: tuple creation timestamp, column emphasis and query dependency DAG.
\par We integrate our lineage computations into the PostgreSQL system via an extension (ProvSQL) and extensive experiments exhibit useful results in terms of accuracy against exact, semiring-based, justifications, especially for the column-based (CV) method which exhibits high precision and high per-level recall. In the experiments, we focus on tuples with \textit{multiple generations} of tuples in their lifelong lineage and analyze them in terms of direct and distant lineage. 
\end{abstract}
\maketitle


\pagestyle{\vldbpagestyle}


%
%
\section{Introduction}
\label{chap:intro}


\subsection*{Data Lineage}
The focus of this work is providing explanations (or justifications) for the existence of tuples in a Database Management System (DBMS, DB). These explanations are also known as \textit{data provenance} \cite{cheney2009provenance}.
Provenance in the literature \cite{Ives2008, Deutch2015} often refers to forms of ``justifying'' the existence of tuples in query results. That is, the provenance context is the state of the database (DB) just before the query execution. The specific type of provenance on which we focus is \textit{lineage} \cite{Cui:2000:TLV:357775.357777}, namely a collection of DB tuples whose existence led to the existence of tuple \textit{t} in a query result.

\subsection*{Distant Lineage}
We take a comprehensive look. We track lineage of tuples \textit{throughout} their existence, while distinguishing between tuples that are inserted explicitly and independently of DB content (these are our ``building blocks'') and tuples that are inserted via a query (or, more generally, that depend on the contents of the DB).
In a real-life setting - tuples that are inserted via a query can be one of the following:
\begin{enumerate*}
    \item A hierarchy of materialized views - where each view can depend both on the DB and on previously defined views.
    \item Tuples that are inserted via a SQL \texttt{INSERT INTO SELECT} statement.
    \item Tuples that are inserted via a SQL \texttt{UPDATE} statement.
    \item A query result with explicitly added data fields, that is added back to some table in the DB. For example, get names of customers retrieved from an \texttt{orders} table, calculate some non-database-resident ``customer profile'' for each customer and insert both into a \texttt{customer\_profile} table for providing future recommendations.
\end{enumerate*}

\subsection*{Approximate Lineage}
As time goes on, provenance information for tuples that are inserted via a query may become complex (e.g., by tracking semiring formulas, as presented in \cite{green2007provenance}, or circuits as presented  in \cite{Deutch2014, Senellart2017}). Thus, our goal is providing ``simple to work with'' and useful approximate lineage (using ML and NLP techniques), while requiring only a \textit{constant} additional space per tuple. This approximate lineage is compared against state of the art ``exact provenance tracking system'' in terms of explainability and maintainability.

\subsection*{An Analyst's Outlook}
\begin{figure}[htb]
    \centering
\begin{tikzpicture}[sibling distance=3em, level distance=3em, node distance=0.2cm,
    every node/.append style={font=\scriptsize}
  ]

\begin{scope}[xshift=2.5cm, local bounding box=analyst]
    \node(e3) [yshift=-2.5cm, person,monitor,mirrored,shirt=blue,minimum size=0.5cm] {};
    \node(e1) [align=left, above right=0.6cm of e3, draw =orange!50,thick, rounded corners=0cm] {
        1. Tuples of interest\\
        2. Topics of interest\\
        3. People of interest\\
        4. Actions of interest
    };
    \node(e2) [align=left, below=1cm of e1, draw =green!50,thick, rounded corners=0cm] {
        1. Reports\\
        2. Alarms
    };
    
    \path[->, font=\tiny] 
        (e1) edge node[sloped, anchor=center, below]{\tiny Examine} (e3) 
        (e3) edge node[sloped, anchor=center, below]{\tiny Produce} (e2);
    
\end{scope}

\begin{scope}[xshift=-2.7cm, local bounding box=data]
    \node(d1) [draw =purple!50, ,thick, shape=rectangle,
    draw, align=center, yshift=-1cm] {Vector sets\\ search structure};
    \node(d2) [below=1cm of d1, draw =purple!50, ,thick, shape=rectangle,
    draw, align=center] {Vector sets-\\ adorned DB};
\end{scope}

\begin{scope}[xshift=0cm, local bounding box=actions]
    \node(a1) [draw =blue!50, shape=rectangle, rounded corners,
    draw, align=center] {Compute ranked lineage\\ candidates for target tuple(s)};
    \node(a2) [below=of a1, draw =blue!50, shape=rectangle, rounded corners,
    draw, align=center] {Verify lineage of target\\ tuple(s) - execute query};
    \node(a3) [below=of a2, draw =blue!50, shape=rectangle, rounded corners,
    draw, align=center] { Specify DB subset\\ for lineage computation};
    \node(a4) [below=of a3, draw =blue!50, shape=rectangle, rounded corners,
    draw, align=center] {Specify TV, CV\\ and hyperparameters};
    \node(a5) [below=of a4, draw =blue!50, shape=rectangle, rounded corners,
    draw, align=center] {Specify optimization\\ methods to apply};
    
\end{scope}

\captionsetup[subfigure]{font=tiny,labelfont=tiny}
\node(a) [below=1.05cm of data] {\parbox{0.2\linewidth}{\subcaption{\tiny Data Environment}\label{subfig:a}}};
\node(b) [below=of actions] {\parbox{0.3\linewidth}{\subcaption{\tiny Repertoire of Actions}\label{subfig:b}}};
\node(c) [right=of b] {\parbox{0.3\linewidth}{\subcaption{\tiny Analyst's working environment}\label{subfig:c}}};

\draw [dotted] ($(actions.north east)!-0.3!(analyst.north west)$) --($(actions.south east)!-0.3!(analyst.south west)$);
\draw [dotted] ($(data.north east)!.5!(actions.north west)$) --($(data.south east)!.5!(actions.south west)$);

\path[<->, font=\tiny] 
        ($(actions.east) + (0.8cm, 0)$) edge node[anchor=center, below]{\tiny Apply} ($(actions.east)$)
        (actions) edge node[anchor=center, below]{\tiny Query} (data);
        
\end{tikzpicture}
    \caption[An Analyst's Working Environment Diagram]{
    An Analyst's Outlook - General Scheme
    }\label{fig:analysts_working_env}

\vspace{-4mm}
\end{figure}
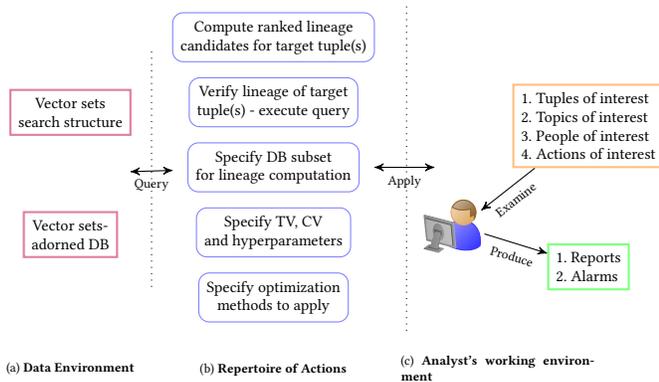
We consider a scenario of an analyst that interacts with a DB with distant lineage tracking capabilities.
Figure \ref{fig:analysts_working_env} shows a general scheme of an analyst's working environment.
The analyst examines tuples/topics/people/actions of interest; he then produces relevant reports with regard to the DB state, or raises alarms for corrupted/suspect data records and operations.
A DB with distant lineage tracking capabilities is adorned with per-tuple or per-column lineage vector sets and a corresponding search structure for those vector sets. Consequently, given a target tuple of interest with its own set(s) of lineage vectors, the analyst can query the vector-sets search structure to compute a list of the top-$k$ ranked (by similarity) lineage candidates for the target tuple.
Finally, given a list of top-$k$ lineage candidates, the analyst may verify the approximate lineage as a sufficient explanation for a target tuple $t$, by re-executing the query that generated $t$ on the obtained lineage candidate tuples.
Depending on the application and the values/sources/other meta-data of the approximate lineage candidates, the analyst produces reports that may be of help for auditing or debugging purposes on a large database system.
\par Consider a real-life example. Susan Black, an analyst for \texttt{Incredible Credit} is presented a case to investigate. The credit rating, as recorded in database record \textit{tvcr}, of one of \texttt{Incredible Credit}’s reviewees, the \texttt{Viola Corp.}, is \texttt{AA} as recorded in the Incredible Credit database (ICDB), despite the fact that \texttt{Viola}’s stock has been plunging over the last six months. The credit rating is the result of a query \textit{QCR} that is executed periodically. \textit{QCR} looks at various data records connected to the \texttt{Viola Corp.}, e.g., earnings, assets, projected growth and additional market-related data, especially in \texttt{Viola}’s market sector, as well as records computed by other queries. Susan queries for the lineage of record \textit{tvcr}. The result is a set \textit{RS} of about 500 records. Susan verifies that indeed the \textit{tvcr} record is in the result of \textit{QCR} applied to \textit{RS}. Skimming through the list of records, Susan notices a record \textit{rerr}, which was used in computing a record \textit{rint} which \textit{QCR} used in computing \textit{tvcr}, in which the assets of the \texttt{Viola Corp.} are estimated at $\$$20B. This looks odd. It turns out that this is a result of an input error and the value should have been $\$$2B. Susan notifies the Business Reporting Dept. at \texttt{Incredible Credit} which fixes the glitch.

\subsection*{Main Contributions}
We present an efficient, word vectors based method, and enhancements thereof, for encoding lifelong lineage, which consumes a constant additional space per tuple.
This work builds upon a preliminary work \cite{DBLP:conf/vldb/LeybovichS20}, and extends it significantly by introducing the concept of verification of approximate lineage, enhancements for distant lineage computations, such as query dependency DAG, and extensive, completely new, distant lineage focused experimentation. Other significant extensions are a better exposition of the system architecture and a description of an analyst's interaction with the system.

\section{Classic Provenance}
\label{chap:classic_provenance}


\subsection{Definition}
Provenance - \textit{source, origin} \cite{prov_dict}. 
In computing, provenance information describes the origins and the history of data within its lifetime. When talking about database management systems, the commonly used term is \textit{data provenance} \cite{cheney2009provenance}. The idea behind data provenance is keeping additional information (meta-data) allowing us to easily answer a number of useful ``meta-questions''.
\par Data provenance helps with providing explanations for the existence of tuples in a query result. The context of these explanations is usually the DB state prior to the query execution.

\subsection{Related Work}\label{sec:related_work}
Over the past 15 years, provenance research has advanced in addressing both theoretical \cite{cheney2009provenance, green2007provenance, Deutch2014} 
and practical \cite{Ives2008, Karvounarakis2010, Deutch2015, Deutch2017, Senellart2018, DBLP:conf/cidr/IvesZHZ19} 
aspects. 
In particular, several different notions of data provenance (\textit{lineage}, \textit{why}, \textit{how} and \textit{where}) were formally defined \cite{Cui:2000:TLV:357775.357777, DBLP:conf/icdt/BunemanKT01,  cheney2009provenance}.
\par A few prior works \cite{approx_lineage, approx_PROX, approx_summary, approx_why_and_why_not} focus on approximate (or summarized) provenance. That is, seeking a compact representation of the provenance, at the possible cost of information loss, in an attempt to deal with the growing size and complexity of exact provenance data in real-life systems.

\par ProvSQL is an open-source project developed by Pierre Senellart et al. \cite{Senellart2018}.
According to \cite{provsql_github}:
``The goal of the ProvSQL project is to add support for (m-)semiring provenance and uncertainty management to PostgreSQL databases, in the form of a PostgreSQL extension/module/plugin.''
ProvSQL uses \textit{circuits}, a compact representation for provenance annotations \cite{Deutch2014, Senellart2017}, which are constructed per-query. A provenance circuit is an inductively built directed acyclic graph (DAG), with the following properties:
\begin{enumerate*}
    \item The leaves contain annotations of tuples from the input DB.
    \item Inner nodes (termed \textit{gates}) represent operators from a particular semiring. 
    \item The edges (termed \textit{wires}) connect nodes to an operator, representing operands of an intermediate calculation.
    \item The sub-DAG under a given node represents the semiring formula for deriving it. 
\end{enumerate*}

\section{Lineage via Embeddings}
\label{chap:lineage_embeddings}

\subsection{Word Embeddings}\label{sec:word_embeddings_intro}
Classic NLP research focuses on understanding the structure of text. For example, building dependency-based parse trees \cite{melcuk1988, klein-manning-2004-corpus} that represent the syntactic structure of a sentence via grammatical relations between its words. These approaches do not account for the \textbf{meaning} of words. \textit{Word embedding} aims to encode meanings of words (i.e., semantics), via low dimension (usually, 200-300) real-valued vectors, which can be used to compute the similarity of words as well as test for analogies \cite{DBLP:conf/naacl/MikolovYZ13}. Two of the most influential methods for computing word embeddings are the \textsc{Word2Vec} family of algorithms, by Mikolov et al. \cite{DBLP:journals/corr/abs-1301-3781, DBLP:conf/nips/MikolovSCCD13} and \textsc{GloVe} by Pennington et al. \cite{pennington2014glove}. Furthermore, applying neural-network (NN) techniques to NLP problems (machine translation \cite{wu2016googles}, named entity recognition \cite{Gillick_2016}, sentiment analysis \cite{Maas:2011:LWV:2002472.2002491} etc.) naturally leads to the representation of words and text as real-valued vectors. 

\subsection{Motivation}\label{sec:motivation_approx}
A few problems become apparent when considering \textit{Distant Lineage} (i.e., indirect, history long, explanations for the existence of tuples in the DB) with traditional and state of the art ``exact provenance tracking'' techniques:
\begin{enumerate*}
    \item Formula based representations may \textit{blow-up} in terms of \textit{space consumption}.
    A naive implementation using semiring polynomials \cite{green2007provenance} requires saving the full provenance polynomial for each tuple, resulting in a massive growth in space consumption (for tuples that are produced by a query and that may depend on result tuples of previous queries).
    
    \item Inductively built representations become very complex over time. Thus, they result in impractical provenance querying time. A naive implementation using circuits \cite{Deutch2014, Senellart2017} would simply keep on constructing provenance circuits as described in section \ref{sec:related_work}. During lineage querying, we may end up with very complex circuits, such that numerous leaves are derived via a circuit of their own (these leaves are tuples that were produced by a previous query and were inserted to the DB). Hence, even if a significant amount of sharing is realized across the provenance circuits - they are inevitably going to blow-up in space consumption. This approach renders keeping and querying the full provenance as impractical and requires limiting heavily the provenance resolution, otherwise (e.g., return a summarized explanation). That is, \textit{querying complex data structures} like circuits is impractical for provenance querying in modern DB systems.
    
    \item Alternatively, if we only want lineage, we could just store with each tuple a set of all the tuples it depends on. This will be cheaper than circuits but still prohibitively expensive. Here too, one could think of circuit-like techniques where tuples that have a subset in common, of tuples in their lineage, could share this subset. But again, this is complex and suffers from similar problems, as discussed above.
    
    \item Complex explanations are not very human-readable. Deutch et al. \cite{Deutch2017} showed how to generate more human-readable explanations - but they are arguably still complex. A ``top-$k$ justifications'' style provenance, which is \textit{simpler} and provides \textit{ranking} of justifications, might be more useful for an analyst in a real-time interaction with the data.
\end{enumerate*}

\par Consider an illustrative example using ProvSQL. Suppose there is a database with two relations: \texttt{Edge(A,B)} - a tuple $(a,b)$ denotes an undirected edge between $a$ and $b$, and \texttt{Path(X,Y)} - a tuple $(x,y)$ denotes a path from node $x$ to node $y$. Suppose that for all $a,b$ that appear in the DB, call them \textit{nodes} and assume there are $\#N$ distinct nodes, \texttt{Edge($a,b$)} exists, i.e., \texttt{Edge} encodes a complete graph. 
Assume $\#E$ is the number of tuples in the \texttt{Edge} table, that is, $\#E = \frac{\#N \cdot (\#N - 1)}{2}$.
Some of the tuples in \texttt{Path} were inserted manually. Others, such as for node 7, were inserted based on the query:
\newcommand{\sqlcolor}[1]{\textcolor{Bittersweet}{#1}}
\newcommand{\fieldcolor}[1]{\textcolor{Fuchsia}{#1}}
\newcommand{\strcolor}[1]{\textcolor{ForestGreen}{#1}}
\newcommand{\numcolor}[1]{\textcolor{TealBlue}{#1}}

\begin{figure}[h]
\vspace{-2mm}
\raggedright
\begin{small}
    \texttt{
    \textcolor{white}{F}\sqlcolor{SELECT DISTINCT} (\fieldcolor{A}, 7) \\
    \textcolor{white}{FF}\sqlcolor{FROM} Edge E$\sb\texttt{1}$, E$\sb\texttt{2}$, E$\sb\texttt{3}$, E$\sb\texttt{4}$, E$\sb\texttt{5}$, E$\sb\texttt{6}$\\
    \textcolor{white}{FF}\sqlcolor{WHERE} E$\sb\texttt{1}$(\fieldcolor{A}, \fieldcolor{B}) \sqlcolor{AND} E$\sb\texttt{2}$(\fieldcolor{B}, \fieldcolor{C}) \sqlcolor{AND} E$\sb\texttt{3}$(\fieldcolor{C}, \fieldcolor{D}) \sqlcolor{AND} \\ 
    \textcolor{white}{FFFFFFFF}E$\sb\texttt{4}$(\fieldcolor{D}, \fieldcolor{E}) \sqlcolor{AND} E$\sb\texttt{5}$(\fieldcolor{E}, \fieldcolor{F}) \sqlcolor{AND} E$\sb\texttt{6}$(\fieldcolor{F}, 7) \sqlcolor{AND} \\
    \textcolor{white}{FFFFFFFF}\fieldcolor{A} $\neq$ \fieldcolor{B} \sqlcolor{AND} \fieldcolor{A} $\neq$ \fieldcolor{C} \sqlcolor{AND} \fieldcolor{A} $\neq$ \fieldcolor{D} \sqlcolor{AND}\\
    \textcolor{white}{FFFFFFFF}\fieldcolor{A} $\neq$ \fieldcolor{E} \sqlcolor{AND} \fieldcolor{A} $\neq$ \fieldcolor{F} \sqlcolor{AND} \fieldcolor{A} $\neq$ 7 \sqlcolor{AND}\\
    \textcolor{white}{FFFFFFFF}\fieldcolor{B} $\neq$ \fieldcolor{C} \sqlcolor{AND} \fieldcolor{B} $\neq$ \fieldcolor{D} \sqlcolor{AND} \fieldcolor{B} $\neq$ \fieldcolor{E} \sqlcolor{AND} \fieldcolor{B} $\neq$ \fieldcolor{F} \sqlcolor{AND} \fieldcolor{B} $\neq$ 7 \sqlcolor{AND}\\
    \textcolor{white}{FFFFFFFF}\fieldcolor{C} $\neq$ \fieldcolor{D} \sqlcolor{AND} \fieldcolor{C} $\neq$ \fieldcolor{E} \sqlcolor{AND} \fieldcolor{C} $\neq$ \fieldcolor{F} \sqlcolor{AND} \fieldcolor{C} $\neq$ 7 \sqlcolor{AND}\\
    \textcolor{white}{FFFFFFFF}\fieldcolor{D} $\neq$ \fieldcolor{E} \sqlcolor{AND} \fieldcolor{D} $\neq$ \fieldcolor{F} \sqlcolor{AND} \fieldcolor{D} $\neq$ 7 \sqlcolor{AND}\\
    \textcolor{white}{FFFFFFFF}\fieldcolor{E} $\neq$ \fieldcolor{F} \sqlcolor{AND} \fieldcolor{E} $\neq$ \fieldcolor{7} \sqlcolor{AND} \fieldcolor{F} $\neq$ 7\\
    }
\end{small}
\vspace{-2mm}
\end{figure}
The result is the tuples \texttt{Path($a,7$)} for all  nodes $a$ in the DB (since the graph is complete, there is a 6-edge length path between every pair of nodes).
Now, the lineage of each such tuple is all the \texttt{Edge} tuples, but the circuit needs to encode the different 6-edge paths separately (it corresponds to a different ordered subset of the set of tuples in \texttt{Edge}, starting with node $a$ and ending with node $7$). For each path from $a$ to $7$, ProvSQL (see section \ref{sec:related_work}) creates a circuit gate encoding the path and 6 wires (for each one of the edges in the path).
There are $(\#N - 2) \cdot (\#N - 3) \cdot (\#N - 4) \cdot (\#N - 5) \cdot (\#N - 6) = O(\#N^5) = O(\#E^{2.5})$
such paths from $a$ to $7$. So, ProvSQL needs to allocate $O(\#E^{2.5})$ gates, each one connected to 7 wires.
In other words, this is very much not linear in the size $O(\#E)$ of the database. Notice that choosing a different query with longer paths only makes matters worse for ProvSQL. On the other hand, the method we present in this paper requires only a constant additional space for each of the query result tuples.

\subsection{Word Vectors Model}\label{sec:latent_wv_model}
The system we propose is based on word vectors models. Such a model is composed of a collection of real-valued vectors, each associated with a relevant DB term. The process of deriving vectors from DB-derived text is called \textit{relational embedding}, which is a very active area of research \cite{10.1145/3318464.3389742, 10.1145/3464509.3464892, arora2020embeddings}.
\subsubsection{Training word vectors} 
As in Bordawekar et al. \cite{DBLP:journals/corr/BordawekarS16}, we train a \textsc{Word2Vec} model \cite{rehurek_lrec} on a corpus that is extracted from the relevant DB.
The training time of word vectors on a DB extracted corpus depends on the size of the corpus, in terms of number of tokens, and on the dimensionality of the trained vectors. For example, on the very small 8.4 MB BFPDB database that we used for experimentation, the training time was ~400 secs, on an 8-core Intel Xeon 5220R machine (no GPU). Various training benchmarks are reported in \cite{DBLP:journals/corr/abs-1301-3781}.
A naive transformation of a DB to unstructured text (a sequence of sentences) can be achieved by simply concatenating the textual representation of the different columns of each tuple into a separate sentence. This approach has several problems \cite{DBLP:journals/corr/BordawekarS16}.
First, when dealing with natural language text, there is an implicit assumption that the semantic influence of a word on a nearby word is inversely proportional to the distance between them. However, not only that a sentence extracted from a tuple does not necessarily correspond to any natural language structure, but, it can be actually thought of as ``a bag of columns''; i.e., the order between different columns in a sentence has no semantic implications.
Additionally, all columns are not the same. That is, some columns-derived terms may hold more semantic importance than others in the same sentence (generated from a tuple). For instance, a primary key column, a foreign key column, or an important domain-specific column (e.g., a manufacturer column in a products table). This implies that in order to derive meaningful embeddings from the trained word vectors, we need to consider inter-column discrimination, during both word vectors training and lineage vectors construction phases.

\par The quality of the word vectors model is crucial to the success of our system. However, optimizing the overall performance should focus not only on the training phase, but also on the way we utilize the trained model. Next, we show one such optimization.
\subsubsection{Sentence embeddings}\label{sec:sentence_embeddings} Extracting sentence embeddings from text has been a well-researched topic in the NLP community over the last seven years. State-of-the-art pre-trained models (e.g., Universal Sentence Encoder \cite{Cer2018UniversalSE, use_github} and BERT \cite{devlin2018bert}) are trained on natural language texts, and thus are not suitable for sentences generated from relational tuples (see discussion above). Hence we train a word embedding model and infer the sentence vectors as a \textit{function of the set of word vectors containing all the words} in a sentence. We average the word vectors for each column separately, and then apply weighted average on the ``column vectors'' (the weight is based on the relative importance of a column, as discussed above). As will be shown, column-based vectors result in significant improvements to lineage encoding.
\section{TV - Per-Tuple Lineage Vectors}
\label{chap:per_tuple_lineage_vectors}

\subsection{Proposed Solution}\label{sec:proposed_solution_approx}
\par We devise a novel approach to lineage tracking, which is based on ML and NLP techniques. The main idea is summarizing, and thereby approximating, the lineage of every tuple with a set of up to $max\_vectors$ (a hyperparameter) vectors. For tuples that are inserted explicitly into the DB, the vectors are obtained using a pre-trained word embeddings model $M$ over the underlying ``text'' of the tuple $t$, applying an intra and inter-column weighted average over the vectors of every word in $t$.
During a query execution process we form the lineage of query result tuples in a similar fashion to that of provenance semirings \cite{green2007provenance}. We extend the + and $\cdot$ operations (see Algorithms \ref{algo:add_approx} and
\ref{algo:mul_approx}, respectively) to generate lineage embeddings, while retaining the ability to propagate information and preserve the representation of lineage via a set of up to $max\_vectors$, constant-size vectors. We obtain lineage embeddings (i.e., vectors) for query output tuples by using this process. These new tuples (and their lineage) may be later inserted into the DB (depending on the specific application).
\removelatexerror
\begin{algorithm2e}
\begin{small}
\SetAlgoLined
\SetKwFunction{ClusterVectorsUsingKMeans}{\fncolor{ClusterVectorsUsingKMeans}}
\SetKwData{pv}{LV$_3$}
\SetKwInOut{HP}{Hyper-Parameters}

\HP{\textit{max\_vectors} - maximum number of vectors per-tuple}
\KwIn{two sets of lineage vectors $LV_1, LV_2$ that represent the lineage of two tuples $t_1$ and $t_2$, respectively}
\KwResult{a new set of lineage vectors \pv, that represents $t_1 + t_2$}
\BlankLine
\BlankLine
 $\pv = LV_1 \cup LV_2$\;
 \If{$|\pv| > max\_vectors$}{
    \cmtcolor{\tcc{
    Divide the vectors into \textit{max\_vectors} groups using k-means and return the centers
    }}
    \pv = \ClusterVectorsUsingKMeans{\paramcolor{$\pv$}}\;
 }
 \caption{Lineage vectors: $\boldsymbol+$ \small{(addition) Algorithm}}\label{algo:add_approx}
 \end{small}
\end{algorithm2e}
\removelatexerror
\begin{algorithm2e}
\begin{small}
\SetAlgoLined
\SetKwFunction{ClusterVectorsUsingKMeans}{\fncolor{ClusterVectorsUsingKMeans}}
\SetKwFunction{Avg}{\fncolor{Avg}}
\SetKwFunction{CartesianProduct}{\fncolor{CartesianProduct}}
\SetKwData{pv}{LV$_3$}
\SetKwInOut{HP}{Hyper-Parameters}

\HP{\textit{max\_vectors} - maximum number of vectors per-tuple}
\KwIn{two sets of lineage vectors $LV_1, LV_2$ that represent the lineage of two tuples $t_1$ and $t_2$, respectively}
\KwResult{a new set of lineage vectors \pv, that represents $t_1 \cdot t_2$}
\BlankLine
\BlankLine
 \pv = $\{\Avg(\paramcolor{v_1, v_2})\mid v_1,v_2 \in \CartesianProduct(\paramcolor{LV_1, LV_2})\}$\;
 \If{$|\pv| > max\_vectors$}{
    \cmtcolor{\tcc{Divide the vectors into \textit{max\_vectors} groups using k-means and return the centers}}
    \pv = \ClusterVectorsUsingKMeans{\paramcolor{$\pv$}}\; 
 }
\caption{Lineage vectors: $\boldsymbol\cdot$ \small{(multiplication) Algorithm}}\label{algo:mul_approx}
\end{small}
\end{algorithm2e}  



\subsubsection{Similarity calculation}\label{sec:similarity_calculation} Given two word vectors, the similarity score is usually the cosine distance between them. In our system, we want to calculate the similarity between the lineage representations of two tuples/columns (see sections \ref{chap:per_tuple_lineage_vectors} and \ref{chap:per_column_lineage_vectors}, respectively); in both cases, the tuple or column, is associated with a \textit{set} of lineage vectors. That is, we need to calculate the similarity between \textit{two sets of vectors}\footnotemark. The ``logic'' behind the following formula is balancing between the ``best pair'' of vectors (in terms of similarity) in the two sets and the similarity between their average vectors:
\begin{equation*}
    \operatorname{sim}(A, B) = \frac{w_{max} \cdot max(ps) + w_{avg} \cdot avg(ps)}
    {w_{max} + w_{avg}}
\end{equation*}
\footnotetext{We note that \cite{DBLP:journals/corr/BordawekarS16, DBLP:journals/corr/abs-1712-07199, DBLP:conf/sigmod/BordawekarS17} have also used various similar methods for measuring similarity between two sets of vectors.}where $ps$ is the set of pair-wise cosine distances between a pair of vectors, one taken from set $A$ and one taken from set $B$. $w_{max}$ and $w_{avg}$ are (user-specified) hyperparameters. $max$ and $avg$ are functions that return the maximum and average values of a collection of numbers, respectively\footnotemark.
\footnotetext{This logic holds for both tuple-based vectors and column-based vectors (i.e., for each column separately).}


\subsubsection{Lineage querying} 
Given a tuple and its lineage embedding vectors, we can calculate the pair-wise similarity against every other tuple in the DB (or a subset, e.g., in a specific table) and return the top $k$ (a parameter) most lineage-similar tuples (these resemble a subset of the lineage \cite{Cui:2000:TLV:357775.357777}). 
There are many algorithms for approximate vector search, e.g., based on LSH \cite{lsh}. Approximate vector search is a very active area of research (see, e.g., \cite{sugawara-etal-2016-approximately}); however, these algorithms are not directly applicable to our method, since we operate on sets of vectors instead of single vectors. We recently devised a method \cite{leybovich2021efficient} that efficiently reduces the set-set similarity problem to vector-vector similarity. This enables using any approximate vector similarity search algorithm to, given a set of vectors, \textit{efficiently} search among sets of vectors for the closest one or $k$-closest ones.


\subsubsection{Intended usage} The intended usage of lineage vectors is as follows:
\begin{enumerate*}
    \item Each manually inserted tuple - has a set consisting of a single tuple vector.
    \item Each tuple in the result of a query - has a set of up to $max\_vectors$ tuple vectors.
    \item When calculating similarity - we always compare between two sets of lineage vectors, by using the formula for $\operatorname{sim}(A, B)$ of section \ref{sec:similarity_calculation}.
\end{enumerate*}
\subsection{Verifying the Lineage}
Given a method that approximates the top-$k$ justifications for the existence of a query result tuple, as the one presented above, there is a simple technique to partially verify the collection of lineage candidates. The idea is applying the query $q$ that generated the tuple to be explained, $t$, to the collection of lineage tuples obtained by our system. If $t$ is output in the result of applying $q$ on these lineage candidates, then we found a sufficiently small (though, non-precise) explanation for $t$. Furthermore, one could utilize an exact provenance tracking application, e.g., ProvSQL \cite{provsql_github} by applying $q$ to the collection of tuples obtained by our technique, to explain $t$ (given that $t$ is ``verified''). This technique relies on the fact that executing $q$ on a small collection of tuples (even with exact provenance tracking) is significantly cheaper than applying the query on the whole, potentially large, DB. In case the computed lineage is insufficient, it may be extended (adjusting parameters).

\subsection{System Life Cycle}
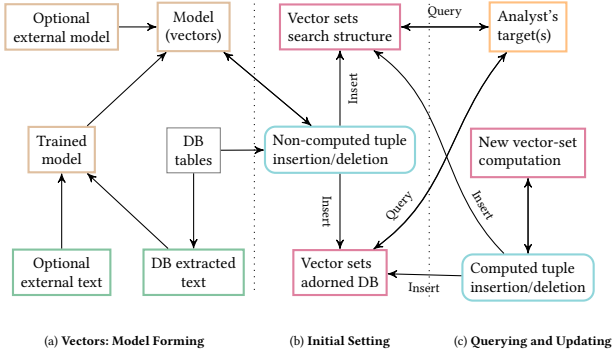
\begin{figure}[htb]
    \centering
\begin{tikzpicture}[sibling distance=3em, level distance=3em, node distance=0.2cm,
    every node/.append style={font=\scriptsize}
  ]

\begin{scope}[xshift=2.5cm, local bounding box=query]
    \node(q1) [align=left, draw =orange!50,thick, rounded corners=0cm] {
        Analyst's\\ target(s)
    };
    \node(q2) [align=left, below=1cm of q1, draw =purple!50,thick, rounded corners=0cm] {
        New vector-set\\
        computation
    };
    \node(q3) [align=left, below=1cm of q2, draw =Aquamarine!50,thick, rounded corners] {
        Computed tuple\\
        insertion/deletion
    };
    
    \path[->, font=\tiny] 
        (q3) edge node[sloped, anchor=center, below]{} (q2) 
        (q2) edge node[sloped, anchor=center, below]{} (q3);
    
\end{scope}

\begin{scope}[xshift=-3.7cm, local bounding box=model]
    \node(m1) [draw =brown!50, ,thick, shape=rectangle,
    draw, align=center] {Optional\\ external model};
    \node(m2) [below=1cm of m1, draw =brown!50, ,thick, shape=rectangle,
    draw, align=center] {Trained\\ model};
    \node(m3) [below=1cm of m2, draw =ForestGreen!50, ,thick, shape=rectangle,
    draw, align=center] {Optional\\ external text};
    \node(m4) [right=0.5cm of m1, draw =brown!50, ,thick, shape=rectangle,
    draw, align=center] {Model\\ (vectors)};
    \node(m5) [below=1cm of m4, draw =black!50, , shape=rectangle,
    draw, align=center, rounded corners=0cm] {DB\\ tables};
    \node(m6) [below=1cm of m5, draw =ForestGreen!50, ,thick, shape=rectangle,
    draw, align=center] {DB extracted\\ text};
    
    \path[->, font=\tiny]
        (m2) edge node[sloped, anchor=center, below]{} (m4)
        (m1) edge node[sloped, anchor=center, below]{} (m4)
        (m5) edge node[sloped, anchor=center, below]{} (m6)
        (m6) edge node[sloped, anchor=center, below]{} (m2) 
        (m3) edge node[sloped, anchor=center, below]{} (m2);
\end{scope}

\begin{scope}[xshift=0cm, local bounding box=init]
        \node(i1) [align=left, draw =purple!50,thick, rounded corners=0cm] {
        Vector sets\\
        search structure
    };
    \node(i2) [align=left, below=1cm of i1, draw =Aquamarine!50,thick, rounded corners] {
        Non-computed tuple\\
        insertion/deletion
    };
    \node(i3) [align=left, below=1cm of i2, draw =purple!50,thick, rounded corners=0cm] {
        Vector sets\\
        adorned DB
    };
    
    \path[->, font=\tiny] 
        (i2) edge node[sloped, anchor=center, below]{\tiny Insert} (i1) 
        (i2) edge node[sloped, anchor=center, below]{\tiny Insert} (i3);
    
\end{scope}

\path[->, font=\tiny]
        (m5) edge node[sloped, anchor=center, below]{} (i2)
        (i2) edge node[sloped, anchor=center, below]{} (m4)
        (m4) edge node[sloped, anchor=center, below]{} (i2)
        
        (q3) edge[in=-35, out=135, pos=0.2] node[sloped, anchor=center, above]{\tiny Insert} (i1) 
        (q3) edge node[sloped, anchor=center, below]{\tiny Insert} (i3)
        
        (i3) edge[out=35, in=230, pos=0.8]  node[sloped, anchor=center, above]{} (q1)
        (q1) edge[in=35, out=230, pos=0.8]  node[sloped, anchor=center, above]{\tiny Query} (i3)
        
        (i1) edge  node[sloped, anchor=center, above]{} (q1)
        (q1) edge  node[sloped, anchor=center, above]{\tiny Query} (i1)
        ;

\captionsetup[subfigure]{font=tiny,labelfont=tiny}
\node(a) [below=of model] {\parbox{0.3\linewidth}{\subcaption{\tiny Vectors: Model Forming}\label{subfig:a}}};
\node(b) [below=of init] {\parbox{0.3\linewidth}{\subcaption{\tiny Initial Setting}\label{subfig:b}}};
\node(c) [right=-0.2cm of b] {\parbox{0.3\linewidth}{\subcaption{\tiny Querying and Updating}\label{subfig:c}}};

\draw [dotted] ($(init.north east)!0.3!(query.north west)$) --($(init.south east)!0.3!(query.south west)$);
\draw [dotted] ($(model.north east)!.5!(init.north west)$) --($(model.south east)!.5!(init.south west)$);

        
\end{tikzpicture}
    \caption[System Life Cycle]{
    System Life Cycle - Chronological Scheme
    }\label{fig:system_life_cycle}

\vspace{-2mm}
\end{figure}
The chronological life cycle of the system is presented in Figure \ref{fig:system_life_cycle}.
First, a word vectors model $M$ is trained on the DB extracted text (and perhaps other texts, obtained from external sources), as detailed in section \ref{sec:latent_wv_model}. Optionally, the trained model may be combined with an external pre-trained model. 
Next, given an initial state of the DB, a set $V_t$ containing a single vector is obtained for each tuple $t$, using the word vectors model $M$ over the underlying text of $t$. $V_t$ is inserted into an efficient vector set search structure \cite{leybovich2021efficient}. Later, explicitly inserted (non-computed) tuples adhere to the same behaviour.
During ongoing operation, vector sets are computed for query results, following the query execution steps (as described in section \ref{sec:proposed_solution_approx}). Similarly, computed vector sets for newly inserted computed tuples may be later inserted into the efficient vector set search structure.
An analyst may use the lineage vector sets to explore the direct/distant lineage of previously inserted computed tuples or current query results.
\section{CV - Per-Column Lineage Vectors}
\label{chap:per_column_lineage_vectors}

\subsection{Rationale} As noted, text obtained from database relations does not behave like natural language text, and a sequence of columns values in a tuple does not usually construct a natural language sentence. Thus, building lineage vectors at the tuple level and comparing tuples on this basis is noisy and potentially lossy in terms of information. Instead, we devise a genetics-inspired approach. Suppose a DB tuple presents the external appearance (i.e., phenotype) of a DB entity, as dictated by the genetic code of the entity. The columns of an entity are its basic properties, i.e., the genes that combine to construct its genetic code. In this setting, querying the direct lineage of a result tuple is analogous to finding its predecessors through DNA tracking. Following the approach presented above, we design an alternative lineage tracking mechanism, that of keeping track of and querying lineage (via embeddings) at the column (gene) level. This way, we manage to better distinguish between the provenance features and the textual characteristics of a tuple.
The rationale behind this method is partially inspired by the construction of ``tuple vectors'' by means of averaging the ``sentence vectors'' of the columns (as discussed in section \ref{sec:sentence_embeddings}).

\subsection{Implementation}\label{sec:cv-implementation}
\begin{enumerate}
    \item Each tuple has a set of (up to \textit{max\_vectors}) \textit{column lineage vectors} per-column, instead of a single set of tuple vectors (per-tuple). Let us denote the set of columns (features, characteristics) for which a tuple $t$ has lineage embeddings as $t.lineage\_columns$.
    \item From now on, we formally denote a column name, using the \textit{full name} notation, as $T.\texttt{Attr}$, such that $T$ is a relation name and \texttt{Attr} is a column name. This is meant to achieve better distinction at the lineage vectors level between tuples that originate from different relations with the same column names. 
    \item Given a tuple $t$, we say that $t.lineage\_columns = $
        \begin{equation*}
            t.native\_columns \cup t.inherited\_columns,
        \end{equation*}
    such that $t.native\_columns$ is the set of data columns of $t$ and $t.inherited\_columns$ is the set of columns (features, characteristics) that $t$ ``inherited'' from its predecessors, but are not necessarily reflected as data columns in $t$.
    Note that the same column name may be in $t$ and inherited from its predecessors as well. This issue will be dealt with shortly.
    In the following examples we use \texttt{A}, \texttt{B}, \texttt{C}, \texttt{D} to represent the full name of a column for brevity. For example, suppose a tuple $t \in T_{AB}$ (a table with only two columns - \texttt{A} and \texttt{B}) has the per-column lineage vectors $CV_t = \{\texttt{A}\!: LV_A,\: \texttt{B}\!: LV_B,\: \texttt{C}\!: LV_C,\: \texttt{D}\!: LV_D\}$ ($CV_t$ is a map of: \textit{$t.lineage\_columns$ $\rightarrow$ sets of lineage vectors}). We say that $t.native\_columns$ = $\{\texttt{A}, \texttt{B}\}$, $t.inherited\_columns$ = $\{\texttt{C}, \texttt{D}\}$ and $t.lineage\_columns$ = $\{\texttt{A}, \texttt{B}, \texttt{C}, \texttt{D}\}$.
    \par\textbf{Note.} From here on, we use the notation $CV_t.columns$ to denote the set of columns in the domain of the map $CV_t$.
    \item When combining lineage embeddings (see Algorithms \ref{algo:add_col_approx},\ref{algo:mul_col_approx}), all calculations are done at the column level.
\removelatexerror
\begin{algorithm2e}
\begin{small}
\SetAlgoLined
\SetKwFunction{ClusterVectorsUsingKMeans}{\fncolor{ClusterVectorsUsingKMeans}}
\SetKwData{pv}{CV$_3$}
\SetKwInOut{HP}{Hyper-Parameters}

\KwIn{two maps of: \textit{columns $\rightarrow$ sets of lineage vectors} $CV_1, CV_2$ that represent the lineage of two tuples $t_1$ and $t_2$, respectively}
\KwResult{a new map of: \textit{columns $\rightarrow$ sets of lineage vectors} \pv, that represents $t_1 + t_2$}
\BlankLine
\BlankLine
 \pv = \{\} $\rightarrow$ \{\}\cmtcolor{\tcp*{an empty map of: \textit{columns $\rightarrow$ sets of lineage vectors}}}
 \ForEach{column $c \in (CV_1.columns \cup CV_2.columns)$}{
    \cmtcolor{\tcc{$CV_i.columns$ is the domain and $CV_i[c]$ is the corresponding \textit{set of lineage vectors} of tuple $t_i$ for the column $c \in CV_i.columns$}}
    \uIf{$c \in (CV_1.columns \cap CV_2.columns)$}{
        $\pv[c]$ = $\paramcolor{CV_1[c]} \:\fncolor{\boldsymbol+}\: \paramcolor{CV_2[c]}$\cmtcolor{\tcp*{see Algorithm \ref{algo:add_approx}}}
    }
    \uElseIf{$c \in CV_1.columns$}{
        $\pv[c]$ = $CV_1[c]$\;
    }
    \Else{
        $\pv[c]$ = $CV_2[c]$\;
    }
 }
\caption{Column Lineage vectors: $\boldsymbol+$ \small{(addition)}}\label{algo:add_col_approx}
\end{small}
 \end{algorithm2e}
\removelatexerror
\begin{algorithm2e}
\begin{small}
\SetAlgoLined
\SetKwFunction{ClusterVectorsUsingKMeans}{\fncolor{ClusterVectorsUsingKMeans}}
\SetKwFunction{Avg}{\fncolor{Avg}}
\SetKwFunction{CartesianProduct}{\fncolor{CartesianProduct}}
\SetKwData{pv}{CV$_3$}
\SetKwInOut{HP}{Hyper-Parameters}

\KwIn{two maps of: \textit{columns $\rightarrow$ sets of lineage vectors} $CV_1, CV_2$ that represent the lineage of two tuples $t_1$ and $t_2$, respectively}
\KwResult{a new map of: \textit{columns $\rightarrow$ sets of lineage vectors} \pv, that represents $t_1 \cdot t_2$}
\BlankLine
\BlankLine
 \pv = \{\} $\rightarrow$ \{\}\cmtcolor{\tcp*{an empty map of: \textit{columns $\rightarrow$ sets of lineage vectors}}}
 \ForEach{column $c \in (CV_1.columns \cup CV_2.columns)$}{
    \cmtcolor{\tcc{$CV_i.columns$ is the domain and $CV_i[c]$ is the corresponding \textit{set of lineage vectors} of tuple $t_i$ for the column $c \in CV_i.columns$}}
    \uIf{$c \in (CV_1.columns \cap CV_2.columns)$}{
        $\pv[c]$ = $\paramcolor{CV_1[c]} \:\fncolor{\boldsymbol\cdot}\: \paramcolor{CV_2[c]}$\cmtcolor{\tcp*{see Algorithm \ref{algo:mul_approx}}}
    }
    \uElseIf{$c \in CV_1.columns$}{
        $\pv[c]$ = $CV_1[c]$\;
    }
    \Else{
        $\pv[c]$ = $CV_2[c]$\;
    }
 }
\caption{Column Lineage vectors: $\boldsymbol\cdot$ \small{(multiplication)}}\label{algo:mul_col_approx}
\end{small}
\end{algorithm2e}
    \item After constructing a new tuple $t \in T$ via a query $q$ and its per-column lineage vectors $CV_t$ (via a series of + and $\cdot$ operations, see Algorithms \ref{algo:add_col_approx} and \ref{algo:mul_col_approx}, respectively) - special care must be taken in constructing lineage vectors per the $native\_columns$ of $t$, which might or might not be inherited from $t$'s predecessors.
    That is, for every column $\texttt{A} \in t.native\_columns$ (\texttt{A} represents the full name of a column for brevity):
    \begin{enumerate}
        \item If $\texttt{A} \notin t.inherited\_columns$ and $t.\texttt{A}$ is set to an existing value from some column $\texttt{A'}$ in the DB, e.g., $q$ = \texttt{(INSERT INTO T SELECT A' FROM ...)} then we set $CV_t[\texttt{A}] = CV_t[\texttt{A'}]$.
        \item If $\texttt{A} \in t.inherited\_columns$ and $t.\texttt{A}$ is set to an existing value from some column $\texttt{A'} \neq \texttt{A}$ in the DB, e.g., $q$ = \texttt{(INSERT INTO T SELECT A' FROM ...)} then we set $CV_t[\texttt{A}] =$ $CV_t[\texttt{A}]\; \cdot \;CV_t[\texttt{A'}]$.
        This way, we incorporate both the existing lineage data and the newly calculated one for the column \texttt{A}.
        \item If $\texttt{A} \notin t.inherited\_columns$ and $t.\texttt{A}$ is set to some constant value, e.g., $q$ = \texttt{(INSERT INTO T SELECT \textit{const} FROM ...)} then we set $CV_t[\texttt{A}] = \{initial\_vector(const)\}$, such that $initial\_vector(const)$ is calculated using the pre-trained word vectors model $M$, as described in section  \ref{sec:proposed_solution_approx}, with the ``textified'' (see section \ref{sec:latent_wv_model}) form of \textit{const} as input data.
        \item If $\texttt{A} \in t.inherited\_columns$ and $t.\texttt{A}$ is set to some constant value, e.g., $q$ = \texttt{(INSERT INTO T SELECT \textit{const} FROM ...)} then we set $CV_t[\texttt{A}] = CV_t[\texttt{A}]\; \cdot \; \{initial\_$ $vector(const)\}$. This way, we incorporate both the existing lineage data and the newly calculated one for the column \texttt{A}.
    \end{enumerate}
    \item When comparing a tuple $t$ (and its lineage embeddings) to a set of other tuples $T'$:
    \begin{enumerate}
        \item If $t'.lineage\_columns \nsubseteq t.lineage\_columns$ ($t' \in T'$) then $t'$ is definitely not a part of the lineage of $t$. Otherwise, all of the ``genes'' (columns) of $t'$ would be reflected in $t$. \label{subsubsec:lineage-columns-filtering}
        \item Ideally, we would have kept all the genes that are affecting the target tuple $t$. However, because of practical reasons (see discussion in \ref{sec:cv_discussion}), sometimes certain genes are dropped.
        Hence, the similarity between $t$ and $t' \in T'$ is averaged over the pair-wise similarities of their respective mutual genes (lineage columns). Notice that without dropping lineage columns, the genes of a potential ancestor $t'$ must be a subset of the genes of $t$.
        For example, say a tuple $t$ that has lineage vectors for columns \texttt{A},\texttt{B},\texttt{C} is compared to another tuple $t'$ that has lineage vectors for columns \texttt{B},\texttt{C},\texttt{D} (here, $t'$ is an arbitrary tuple, which is not necessarily in the lineage of $t$):
        \begin{equation*}
            \operatorname{sim}(t, t') = avg(\{\operatorname{sim}(t.\texttt{B}, t'.\texttt{B}), 
            \operatorname{sim}(t.\texttt{C}, t'.\texttt{C})\})
        \end{equation*}
        where $avg$ is a function that returns the average of a collection of numbers. $\operatorname{sim}(t.\texttt{B}, t'.\texttt{B})$ is the calculated similarity between the lineage vectors for column \texttt{B} of $t$ and $t'$ (similarly for column \texttt{C}). Observe that \texttt{A} and \texttt{D} are not mutual ``genes'' of the tuples $t$ and $t'$, and, thus they do not hold lineage information that is valuable to the similarity calculation.
    \end{enumerate}

\end{enumerate}

\begin{runexample}
    Next, we show examples of the addition and multiplication column lineage vectors constructions (Algorithms \ref{algo:add_col_approx} and \ref{algo:mul_col_approx}, respectively).
    Suppose we have two tuples $t_1, t_2$ with respective maps of \textit{columns $\rightarrow$ sets of lineage vectors} $CV_1, CV_2$, such that:
    \begin{equation*}
        CV_1 = \{\texttt{A}\!: LV_A,\: \texttt{B}\!: LV_{B_1}\},\: CV_2 = \{\texttt{B}\!: LV_{B_2},\: \texttt{C}\!: LV_C\}
    \end{equation*}
    $\texttt{A}, \texttt{B}, \texttt{C}$ are full column names and $LV_A, LV_{B_1}, LV_{B_2}, LV_C$ are sets of lineage vectors.
\end{runexample}
\begin{example-withrun}\label{example:algo_add_col}
    Let us follow the construction of $CV_3$, which represents the column lineage embeddings of $t_1 + t_2$ using Algorithm \ref{algo:add_col_approx} (corresponds to alternative use of data, i.e., OR in the query):
    \begin{enumerate}
        \item $CV_3 = \{\} \rightarrow \{\}$
        \item $CV_1.columns \,\cup\, CV_2.columns = \{\texttt{A}, \texttt{B}\} \cup \{\texttt{B}, \texttt{C}\} = \{\texttt{A}, \texttt{B}, \texttt{C}\}$
        \item $\texttt{A} \in CV_1.columns \wedge \texttt{A} \notin CV_2.columns $\\ $\Rightarrow CV_3[\texttt{A}] = CV_1[\texttt{A}] = LV_A$
        \item $\texttt{B} \in (CV_1.columns \cap CV_2.columns) $\\ $\Rightarrow CV_3[\texttt{B}] = CV_1[\texttt{B}] + CV_2[\texttt{B}] = LV_{B_1} + LV_{B_2}$
        \item $\texttt{C} \notin CV_1.columns \wedge \texttt{C} \in CV_2.columns $\\ $\Rightarrow CV_3[\texttt{C}] = CV_2[\texttt{C}] = LV_C$
        \item $CV_3 = \{\texttt{A}\!: LV_A,\: \texttt{B}\!: LV_{B_1} + LV_{B_2},\: \texttt{C}\!: LV_C\}$
    \end{enumerate}
\end{example-withrun}
\begin{example-withrun}
    Let us follow the construction of $CV_3$, that represents the column lineage embeddings of $t_1 \cdot t_2$ using Algorithm \ref{algo:mul_col_approx} (corresponds to joint use of data, i.e., AND in the query):
    \begin{enumerate}
        \item $CV_3 = \{\} \rightarrow \{\}$
        \item $CV_1.columns \,\cup\, CV_2.columns = \{\texttt{A}, \texttt{B}\} \cup \{\texttt{B}, \texttt{C}\} = \{\texttt{A}, \texttt{B}, \texttt{C}\}$
        \item $\texttt{A} \in CV_1.columns \wedge \texttt{A} \notin CV_2.columns $\\ $\Rightarrow CV_3[\texttt{A}] = CV_1[\texttt{A}] = LV_A$
        \item $\texttt{B} \in (CV_1.columns \cap CV_2.columns) $\\ $\Rightarrow CV_3[\texttt{B}] = CV_1[\texttt{B}] \cdot CV_2[\texttt{B}] = LV_{B_1} \cdot LV_{B_2}$
        \item $\texttt{C} \notin CV_1.columns \wedge \texttt{C} \in CV_2.columns $\\ $\Rightarrow CV_3[\texttt{C}] = CV_2[\texttt{C}] = LV_C$
        \item $CV_3 = \{\texttt{A}\!: LV_A,\: \texttt{B}\!: LV_{B_1} \cdot LV_{B_2},\: \texttt{C}\!: LV_C\}$
    \end{enumerate}
\end{example-withrun}

\subsubsection{Dropping lineage columns}\label{sec:cv_discussion}
$t.lineage\_columns$ (for a tuple $t$) might get large and include (almost) all the columns in the DB. This can render this solution impractical, since large modern systems may operate with hundreds and even more columns across the DB. There are various practical techniques that can serve to limit the number of lineage columns per tuple:
\begin{enumerate*}
    \item Set a bound \textit{b} on the number of $lineage\_columns$ (a hyperparameter).
    \item Give \textit{native priority}. That is, prefer keeping $native\_columns$ and cutting-off $inherited\_columns$, whose influence on \textit{this} tuple is more remote, when the bound is reached (see example \ref{example:generational_priority} below).
    \item Remove the relation name prefix from a column name, i.e., consider all $T.\texttt{Attr}$ as simply \texttt{Attr}. This might result in a loss of information, but hopefully not too harmful to the overall performance of the method.
\end{enumerate*}
Observe that the assumption in \ref{subsubsec:lineage-columns-filtering} of section \ref{sec:cv-implementation} breaks down under this lineage column cut-off technique. We can either live with the false negatives or suggest filtering only if the containment rate of $t'.lineage\_columns$ in $t.lineage\_columns$ is below some threshold.

\begin{example-withrun}\label{example:generational_priority}
    Let us follow the construction of $CV_3$, that represents the column lineage embeddings of $t_3 = t_1 + t_2$, such that the result is a new tuple $t_3 \in T_{AB}$ (a table with only two columns - \texttt{A} and \texttt{B}). In addition, we use a bound of $b = 2$ on the number of retained $lineage\_columns$ and implement native priority:
    \begin{enumerate}
        \item (see exapmle \ref{example:algo_add_col})\\ $CV_3 = \{\texttt{A}\!: LV_A,\: \texttt{B}\!: LV_{B_1} + LV_{B_2},\: \texttt{C}\!: LV_C\}$\\ $\Rightarrow t_3.lineage\_columns = \{\texttt{A}, \texttt{B}, \texttt{C}\}$
        \item $t_3 \in T_{AB}$ $\Rightarrow t_3.native\_columns = \{\texttt{A}, \texttt{B}\}$, \\$t_3.inherited\_columns = \{\texttt{C}\}$
        \item $|t_3.lineage\_columns| = 3 > 2 = b$
        \item $CV_3 = \{\texttt{A}\!: LV_A,\: \texttt{B}\!: LV_{B_1} + LV_{B_2}$\}\\ (the \textit{inherited\_column} \texttt{C} was eliminated)
    \end{enumerate}
\end{example-withrun}

\section{Improving Lineage Embeddings}
\label{chap:improving_lineage_embeddings}

\subsection{Tuple Creation Timestamp}
Filtering out non-lineage tuples by a tuple's creation timestamp is a good practice, and can help with filtering out very similar but non-lineage tuples, which might not be detected by the other methods. It is especially important for querying distant lineage (when the DB has existed for a sufficiently long time, so that a tuple's creation timestamp becomes a strong distinguishing factor).


\subsection{Query-Dependent Column Weighting}\label{sec:query-dependent-weighting}
\subsubsection{Rationale} When analyzing lineage embeddings of query-result tuples against other tuples in the DB, we would like to emphasize certain \textit{columns of interest}. These columns can be derived from the structure of the query. For example, if a query asks for distinct manufacturer names of products that contain soda, then intuitively, we want the \texttt{manufacturer} and \texttt{ingredients} columns (of the \texttt{products} table) to have more influence on the respective lineage embeddings than other columns that are not mentioned in the query.

\subsubsection{Implementation} 
\begin{enumerate}
    \item Given a query $q$, we parse $q$ to retrieve the columns of interest and save them as additional meta information for every tuple in the result of $q$.
    \item When comparing a tuple from the result of $q$ with another tuple from the DB, we compare between respective column vectors and calculate a weighted average of the similarities while prioritizing the columns of interest. For example, say a tuple $t$ that has lineage vectors for columns \texttt{A},\texttt{B} was created by a query $q$, such that $q.cols\_of\_interest$ = \{\texttt{B}\}; and we compare $t$ with another tuple $t'$ that has lineage vectors for columns \texttt{A},\texttt{B}:
    \begin{equation*}
        \operatorname{sim}(t, t') = \frac{w_\texttt{A}*\operatorname{sim}(t.\texttt{A}, t'.\texttt{A}) +
        w_\texttt{B}*\operatorname{sim}(t.\texttt{B}, t'.\texttt{B})}{w_\texttt{A} + w_\texttt{B}}
    \end{equation*}
    where $w_\texttt{A}$ and $w_\texttt{B}$ are respective column weights such that $w_\texttt{B} > w_\texttt{A}$ as column \texttt{B} is mentioned in the query. $\operatorname{sim}(t.\texttt{A}, t'.\texttt{A})$ is the calculated similarity between the lineage vectors for column \texttt{A} of $t$ and $t'$ (similarly for column \texttt{B}). Observe that although column \texttt{A} is not a column of interest in the example, it is a mutual ``gene'' of tuples $t$ and $t'$, and, thus it still holds lineage information that is valuable to the similarity calculation.
\end{enumerate}

\subsubsection{Note} This technique is applicable mainly for \textit{direct} lineage. In the context of distant lineage we can use it to get ``immediate contributors'' but also take into account lineage tuples identified without this technique (they will naturally tend to be ranked lower).

\subsection{Weighting with Query Dependency DAG}
\subsubsection{Rationale} We would like to distinguish between very similar/nearly identical tuples (text-wise) during both direct and distant lineage querying. 
Additionally, we want to enhance the natural ranking of lineage tuples by amplifying the ``generational gaps''. The idea is that tuples from earlier generations in the distant lineage tree structure (of a query-inserted tuple) are to be assigned a lower similarity score during distant lineage querying. 
The technique is keeping track of \textit{dependencies between queries} in a DAG structure and weighting the similarity scores of query-inserted tuples (during lineage querying) inversely proportional to the distance, between their inserting queries to the query that computed the explained tuple, in the query dependency DAG. We say that a query $q$ \textit{depends on} a query $p$ if a tuple $t_p$ that was created by $p$ was meaningfully involved in the evaluation of $q$ (i.e., $t_p$ is in the distant lineage of some result tuple $t_q$ in the result of $q$).
This approach enables us to give lower similarity scores to tuples that were not involved meaningfully in the evaluation of a query (directly or remotely) and amplify the natural ranking of the results in terms of query creation dependencies.

\subsubsection{Implementation}
\begin{enumerate}
    \item If there is a tuple $t_p$ that is involved meaningfully in the evaluation of a query $q$: we say that $q$ \textit{depends on} the query $p$ that inserted $t_p$ into the DB.
    \item Initialize an empty query dependency DAG $G = (V, E)$ with a maximum number of nodes $S$ and a maximum height $H$ (hyperparameters), such that $V$ is a set of queries and $E = \{(q, p) \in V\! \times\!V\; |\; q\: depends\: on\: p\}$. 
    \item When comparing a tuple $t_q$ (and its lineage embeddings) that was created by a query $q$ (as recorded with this tuple) to another tuple $t_p$ in the DB, we can \textbf{replace} similarity calculations with the following steps:
    \begin{enumerate}
        \item Denote the sub-tree of $q$ in $G$ (i.e., rooted at $q$) as $G_q = (V_q, E_q)$.
        \item Denote the query that created the tuple $t_p$ (as recorded with this tuple) as $p$.
        \item If $p \in V_q$ then multiply the similarity for $t_p$ by $w_d \leq 1$, which is inversely proportional to the distance $d$ from $q$ to $p$ in $G_q$. One possible implementation for the distance-dependent weighting is $w_d = max\{\frac{1}{2}, 1 - \frac{(d - 1)}{10}\}$, such that $w_1 = 1$, $w_2 = \frac{9}{10}$, etc., and $\forall 1\leq d\leq H: w_d \geq \frac{1}{2}$. Other implementations exist as well. 
        \item If $p \notin V_q$ then multiply the similarity for $t_p$ by $0 <$ $w_{outsider} <$ $w_{d=H}$ (a hyperparameter). Note that since we defined limits on the maximum number of nodes $S$ and maximum height $H$, some nodes will need to be removed when those limits are reached (see details below). Thus, this method might start producing false negatives at some point; and to not lose those tuples completely, we choose to multiply their similarity by some small, non-zero constant, instead of plainly filtering them out.
    \end{enumerate}
    \item We emphasize a lineage tracking system with constant additional space per tuple. The query dependency DAG adheres to this philosophy and is limited in size by the hyperparameters $S$ and $H$. Several design choices were made to enforce these limits, while maintaining effectiveness.
\end{enumerate}

\section{Experimental Evaluation}
\label{chap:experimental_evaluation}

In this section we first establish an explanation quality measure, in terms of precision and per-level recall. We then present and analyze results of our approximate lineage system experimentation against an ``exact provenance tracking system'' - ProvSQL.

\subsection{Experimental Setup}\label{sec:evaluation-setup}
\par\textbf{Note.} \textit{max\_vectors\_num} (see section \ref{sec:proposed_solution_approx}) was chosen manually to be 4. The choice of hyperparameters, and, specifically, \textit{max\_vectors\_num} - is the subject of ongoing research.\\

\subsubsection{Precision Calculation}\label{sec:precision}
As stated in section \ref{sec:proposed_solution_approx}, we expect our explanations to approximate the exact lineage of query result tuples. 
Thus, in order to test the aforementioned algorithms and implementation, we devised an explanation quality measure for explaining a single tuple $t$, where $k$ is a parameter: 
\begin{equation*}
    \operatorname{Precision}(t, k) = \frac{|ApproxLineage(t, k) \cap ExactLineage(t)|}{|ApproxLineage(t, k)|}
\end{equation*}
where $ApproxLineage(t, k)$ is the set of the top $k$ (by lineage similarity) tuples, returned as explanations (i.e., approximate lineage) for $t$ by our system. $ExactLineage(t)$ is the set of tuples that comprise the exact \textbf{distant} lineage of $t$, it can be calculated recursively from the semiring polynomial that is returned by the ProvSQL system for $t$.
For example, if the direct lineage of $t_4$ and $t_5$ are the sets $\{t_1, t_2, t_3\}$ and $\{t_1\}$, respectively, and the direct lineage of a tuple $t_6$ is the set $\{t_4, t_5\}$ then the total distant lineage for the tuple $t_6$ is $\{t_1, t_2, t_3, t_4, t_5\}$.
The parameter $k$ is set by us in each experiment, that is, it is a parameter of the experiment. 
$\operatorname{Precision(t, k)}$ is tunable via the parameter $k$, i.e., when $k$ is small we test for precision of the ``top'' explanations we found.
\par Classically, we are usually interested in precision and recall (e.g., in traditional statistics and classification problems). However, here the situation is slightly different, what we really are interested in is to assess the quality of our explanations by measuring ``how many of the top $k$ (by lineage similarity) tuples are actually part of the exact lineage?''. By contrast, traditional recall does not seem to be a meaningful metric in our case, as many query-result tuples might have long histories. Hence, a top-$k$ justifications result is preferred over ``returning all the correct lineage tuples'', and this is what we measure.

\subsubsection{Distant Lineage: Per-Level Recall}\label{sec:recall}
    In order to assess the quality of a distant lineage answer in a more insightful way, in the experiments we save the ``exact distant lineage'' in a hierarchical structure, and analyze it per-level. Consider a list $L$ of sets per-tuple, s.t. $L[i]$ is the set of all lineage tuples at the $i^{th}$ derivation depth level. Note that $i$ starts at $0$, s.t. the $0^{th}$ lineage level is a set containing the target tuple only. For example, if the direct lineage of $t_4$ and $t_5$ are the sets $\{t_1, t_2, t_3\}$ and $\{t_1\}$, respectively, and the direct lineage of a tuple $t_6$ is the set $\{t_3, t_4, t_5\}$ (note that $t_3$ appears in both the $1^{st}$ and the $2^{nd}$ lineage levels of $t_6$, in this example) then the hierarchical lineage DAG for the tuple $t_6$, and the hierarchical list structure that represents it, look as follows: \\\\
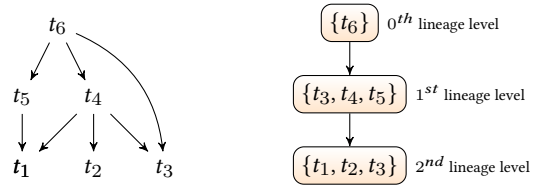
\begin{tikzpicture}[sibling distance=3em, level distance=3em,
  edge from parent/.style={->,draw},
  every edge/.style={->,draw}
  ]]
  \node(t_6) {$t_6$}
    child { node {$t_5$} 
      child { node {$t_1$} } 
      }
    child { node {$t_4$}
      child { node {$t_1$} }
      child { node {$t_2$} }
      child { node(t_3) {$t_3$} }
      };
      
    \draw (t_6) edge[bend left] (t_3);
\end{tikzpicture} \hspace{35pt}
\begin{tikzpicture}[sibling distance=3em, level distance=3em,
  every node/.style = {shape=rectangle, rounded corners,
    draw, align=center,
    top color=white, bottom color=white},
  edge from parent/.style={->,draw}
  ]]
  \node [bottom color =orange!20, label={right:{\scriptsize $0^{th}$ lineage level}}] {$\{t_6\}$}
    child { node [bottom color =orange!20, label={right:{\scriptsize $1^{st}$ lineage level}}] {$\{t_3, t_4, t_5\}$} 
      child { node [bottom color =orange!20, label={right:{\scriptsize $2^{nd}$ lineage level}}] {$\{t_1, t_2, t_3\}$} } 
      };
\end{tikzpicture} \\\\
Here, each rectangle is an entry in the list $L$.
\par We expect our explanations to have a ``natural ranking'' property in terms of lineage levels. That is, we expect the similarity between a target tuple and tuples in its distant lineage to be inversely related to the distance between them in the hierarchical lineage DAG structure. Thus, we devise an explanation quality measure $\operatorname{Recall}(t, i)$ for the explanation of a single tuple $t$ and its lineage level $i$. Let $D(t, i)$ be the number of unique returned tuples in the exact lineage of $t$ up until the $i^{th}$ level (including); formally, $D(t, i) = |\bigcup\limits_{j=1}^{i} L_t[j]|$. Define
\begin{equation*}
    \operatorname{Recall}(t, i) = \frac{|ApproxLineage(t, D(t, i)) \cap L_t[i])|}{|L_t[i]|}
\end{equation*}
where $ApproxLineage(t, D(t, i))$ is the set of the top $D(t, i)$ (by lineage similarity) tuples, returned as explanations (i.e., approximate lineage) for $t$ by our system.
$L_t$ is the hierarchical list structure for $t$, as defined above, i.e., $L_t[i]$ is the set of all lineage tuples at the $i^{th}$ derivation depth level of $t$.

\subsection{Experimenting with Direct and Multi-Generation Lineage}\label{sec:advanced_experiments}
In this section we focus on tuples with multiple generations in their lifelong lineage history and analyze them in terms of direct and \textbf{distant} lineage.
We assess the performance of our system quantitatively using precision and per-level recall (see sections \ref{sec:precision} and \ref{sec:recall}, respectively) and qualitatively by observing the ``top-$k$'' returned lineage tuples (by similarity).
\par Note that
suggested improvements (section \ref{chap:improving_lineage_embeddings}) such as query-dependent column weighting, tuple creation timestamp and weighting with query dependency DAG are used (in the following experiments) for both the Tuple Vectors and the Column Vectors approximate lineage computation methods.

\subsubsection{The BFPDB Dataset}
The USDA Branded Food Products Database (BFPDB) \cite{usda_bfpd-dataset} is the result of a Public-Private Partnership, whose goal is to enhance public health and the sharing of open data by complementing USDA Food Composition Databases with nutrient composition of branded foods and private label data, provided by the food industry. Among others, the dataset includes three tables:
    \begin{enumerate*}
        \item \texttt{products} - contains basic information on branded products (manufacturer, name, ingredients).
        \item \texttt{nutrients} - each tuple contains all nutrient value information about some product.
        \item \texttt{serving\_size} - each tuple contains information about a product's serving size (as well as unit of measurement) and food preparation state.
    \end{enumerate*}
    Each product in the BFPDB dataset has a unique identifier, \texttt{ndb\_no}.

\subsubsection{A Hierarchy of Materialized Views}
We want to simulate a DBMS that contains a significant portion of tuples that depend on the contents of the DB, as a \textit{platform} for testing and analyzing distant lineage. Thus, we built a hierarchy of ``materialized views'', such that each ``view'' is a table, constructed from tuples that were generated by a single query.
In particular, we detail the following materialized views (the text colors are intended as a visual aid, such that each lineage generation has a brighter color, the farther it is from the base relations):
\begin{enumerate*}
    \item \textcolor{Orchid}{\texttt{sugars}} was directly created from tuples of the tables \textcolor{Violet}{\texttt{products}} and \textcolor{Violet}{\texttt{nutrients}}. It contains all the products-related information from the \texttt{products} table, for products that have a \textit{sugars} nutrient-related information in the \texttt{nutrients} table.
    \item \textcolor{CarnationPink}{$\texttt{exp}\sb\texttt{2}$} was directly created from tuples of the materialized views \textcolor{Orchid}{\texttt{unprepared}} and \textcolor{Orchid}{\texttt{protein}}. It contains distinct manufacturers of products that have \textit{water} as an ingredient, and contain \textit{protein} nutrient information; also, these manufacturers produce \textit{unprepared} products.
    \item \textcolor{CarnationPink}{$\texttt{exp}\sb\texttt{3}$} was directly created from tuples of the materialized views \textcolor{Orchid}{\texttt{prepared}}, \textcolor{Orchid}{\texttt{unprepared}} and \textcolor{Orchid}{\texttt{protein}}. It contains distinct manufacturers of products that have \textit{water} and \textit{sugar} as an ingredient, and contain \textit{protein} nutrient information; also, these manufacturers produce \textit{prepared} and \textit{unprepared} products.
    \item \textcolor{Salmon}{$\texttt{exp}\sb\texttt{4}$} was directly created from tuples of the materialized views \textcolor{CarnationPink}{$\texttt{exp}\sb\texttt{2}$} and \textcolor{Orchid}{\texttt{readytodrink}}. It contains distinct names of \texttt{readytodrink} products that contain \textit{mango} in their name, and are produced by manufacturers from the $\texttt{exp}\sb\texttt{2}$ materialized view.
\end{enumerate*}
\par As stated earlier, we consider those tables that do not depend on the contents of the DB when these tables are created, and thereafter, as \textit{base tables}. Again, note that tuples can be manually inserted to and deleted from base tables, but, not in a programmatic manner over the DB, e.g., via a SQL \texttt{INSERT} statement that is based on the DB contents. In our case, the base tables are: \texttt{products}, \texttt{nutrients} and \texttt{serving\_size}. Note that each materialized view depends \textit{directly} on tuples from the base tables (as defined above) or on tuples from other previously constructed materialized views.


\subsubsection{Experiments}
\begin{runexperiment}
We test queries on a subset of the BFPDB dataset, that consists of all the tables and materialized views discussed above. In the experiments we mimic an analyst's interaction with the data by comparing the approximate lineage vectors of a target tuple with a ``heterogeneous'' group of tuples (e.g., all related base tables or all related materialized views) and ranking all the tuples among the group according to similarity scores.
\end{runexperiment}

\begin{experiment-withrun}\label{advanced-experiment:1}
We ask for distinct \texttt{manufacturers} that appear in the $\texttt{exp}\sb\texttt{3}$ materialized view, and produce products that have \textit{salt} as an ingredient:

\begin{figure}[h]
\vspace{-2mm}
\raggedright
\begin{small}
    \texttt{
    \textcolor{white}{F}\sqlcolor{SELECT} p.\fieldcolor{manufacturer} \\
    \textcolor{white}{FF}\sqlcolor{FROM} exp$\sb\texttt{3}$, products p\\
    \textcolor{white}{FF}\sqlcolor{WHERE} exp$\sb\texttt{3}$.\fieldcolor{manufacturer} = p.\fieldcolor{manufacturer} \\ 
    \textcolor{white}{FF}\sqlcolor{AND POSITION}(\strcolor{'salt'} \sqlcolor{IN} p.\fieldcolor{ingredients}) > \numcolor{0} \\
    \textcolor{white}{FF}\sqlcolor{GROUP BY} p.\fieldcolor{manufacturer}
    }
\end{small}
\vspace{-2mm}
\end{figure}
\begin{table}
\centering
\begin{adjustbox}{width=\columnwidth,center}
\pgfplotstabletypeset[
    color cells={min=0.32,max=1.0}, format index, format stats-table1-exp1, format manufacturer, format method,
    col sep=comma,
    /pgfplots/colormap={orange}{
        rgb255(0cm)=(255,245,235);
        rgb255(1cm)=(253,141,60)},
    assign column name/.code=\pgfkeyssetvalue{/pgfplots/table/column
        name}{{\textbf{#1}}},
]{
    
    index,  Provenance Method,       manufacturer, Total, Length-L1, Length-L2, Length-L3,  {1.00$^{Top}$},            {0.75$^{Top}$},        {0.50$^{Top}$},        {0.25$^{Top}$}, Recall-L1, Recall-L3  
    0,      Tuple Vectors,   red gold, 160, 63, 0, 160,          0.600 ,        0.636 ,     0.642 ,       0.585, 0.333, 0.6
    1,      Column Vectors,   red gold, 160, 63, 0, 160,          0.744 ,        0.901 ,     0.988 ,       1.0, 0.778, 0.744
}
\end{adjustbox}
\textbf{\caption{\label{tab:experiment1-base-tables:precision-and-recall}
                Experiment \ref{advanced-experiment:1} \textit{red gold} vs. related base tables (\texttt{products}, \texttt{nutrients} and \texttt{serving\_size}).
                }
        }
\vspace{-4mm}
\end{table}
\begin{table}
\centering
\begin{adjustbox}{width=\columnwidth,center}
\pgfplotstabletypeset[
    color cells={min=0.7,max=0.9, textcolor=white}, format index, format analyst, format Lineage,
    col sep=comma,
    /pgfplots/colormap={purple}{
        rgb255(0cm)=(239,237,245);
        rgb255(1cm)=(106,81,163)},
    assign column name/.code=\pgfkeyssetvalue{/pgfplots/table/column
        name}{{\textbf{#1}}},
]{
    index,     tuple-id, related-table,      {Similarity},      lineage-level,  Lineage 
    0,     $id_1$,    nutrients,    0.864,    {3}          ,\textcolor{Green}{Yes}
    1,     $id_2$,    products,     0.861,    {1, 3}          ,\textcolor{Green}{Yes}
    2,     $id_3$,    products,    0.858,    {1, 3}          ,\textcolor{Green}{Yes}
    3,     $id_4$,    products,    0.857,    {1, 3}          ,\textcolor{Green}{Yes}
    4,     $id_5$,    products,    0.856,    {1, 3}          ,\textcolor{Green}{Yes}
    5,     $id_6$,    products,    0.856,    {1, 3}          ,\textcolor{Green}{Yes}
    $\vdots$,     ,    $\vdots$,    ,              $\vdots$,
    16,     $id_{17}$,    products,    0.847,    {1, 3}          ,\textcolor{Green}{Yes}
    17,     $id_{18}$,    products,    0.847,    {1, 3}          ,\textcolor{Green}{Yes}
    18,     $id_{19}$,    products,    0.847,    {3}          ,\textcolor{Green}{Yes}
    19,     $id_{20}$,    products,    0.846,    {3}          ,\textcolor{Green}{Yes}

}
\end{adjustbox}
\textbf{\caption{\label{tab:experiment1-base-tables:analyst-cv}
                Experiment \ref{advanced-experiment:1} \textit{red gold} vs. related \textbf{base tables} with \textbf{Column Vectors} - an analyst's view.
                }
        }
\vspace{-7mm}
\end{table}
Table \ref{tab:experiment1-base-tables:precision-and-recall} presents lineage related statistics, collected when computing the approximate lineage of the single result tuple in the output of this query (\textit{red gold}) against all the tuples from the related \textbf{base tables}: \texttt{products}, \texttt{nutrients} and \texttt{serving\_size}, which contain $29,322$ tuples in total. Tables \ref{tab:experiment1-base-tables:analyst-cv} and \ref{tab:experiment1-base-tables:analyst-tv} present the top 20 tuples from the related base tables, ranked by similarity, using the CV and TV methods, respectively. We make the following observations by analyzing the results:
\begin{enumerate*}
    \item The lineage sizes of the different levels (1-3) sum up to more than the total distant lineage size ($63 + 0 + 160 > 160$). In this case, it means that all 63 tuples from the $1^{st}$ lineage level (which contains tuples from the \texttt{products} relation) are also a part of the $3^{rd}$ lineage level (which contains tuples from the \texttt{products}, \texttt{nutrients} and \texttt{serving\_size} relations). From analyzing the query we conclude that these are the 63 tuples from the \texttt{products} table that affect the query result directly (via the \texttt{FROM} clause) and indirectly (via tuples from $\texttt{exp}\sb\texttt{3}$).
    \item The Column Vectors (CV) method demonstrates overall superiority, compared to the Tuple Vectors (TV) method. Moreover, the  CV method has a precision of 0.9 for the top $0.75\cdot160 = 120$ tuples in the approximate lineage, and 0.74 for the top 160 tuples. That is, most of the errors are produced for the lower ranked tuples (ranked 121, ..., 160).
    \item The CV method exhibits relatively high L[1] and L[3] recall scores (see Table \ref{tab:experiment1-base-tables:precision-and-recall}), evidenced by observing the Lineage Level(s) column in Table \ref{tab:experiment1-base-tables:analyst-cv}. That is, not only that all the top-20 tuples by similarity are really a part of the distant lineage, but also, the ranking preserves a level-based bias (most of the top-20 are in the $1^{st}$ lineage level). By contrast, the TV method is less impressive on this front, as is evidenced by the \textcolor{Red}{No} results in Table \ref{tab:experiment1-base-tables:analyst-tv}.
    \item Note that a random choice of the top 160 lineage tuples would have yielded a $\frac{160}{29,322} \approx 0.006$ precision score, which is several orders of magnitude worse than the scores of the CV and TV methods. 
\end{enumerate*}
\begin{table}
\centering
\begin{adjustbox}{width=\columnwidth,center}
\pgfplotstabletypeset[
    color cells={min=0.7,max=0.8, textcolor=white}, format index, format analyst, format Lineage,
    col sep=comma,
    /pgfplots/colormap={purple}{
        rgb255(0cm)=(239,237,245);
        rgb255(1cm)=(106,81,163)},
    assign column name/.code=\pgfkeyssetvalue{/pgfplots/table/column
        name}{{\textbf{#1}}},
]{
    index,     tuple-id, related-table,      {Similarity},      lineage-level,  Lineage 
    0,     $id_1$,    products,    0.782,    {1, 3}          ,\textcolor{Green}{Yes}
    1,     $id_2$,    products,     0.779,    {1, 3}          ,\textcolor{Green}{Yes}
    2,     $id_3$,    products,    0.778,    {1, 3}          ,\textcolor{Green}{Yes}
    3,     $id_4$,    products,    0.773,    {1, 3}          ,\textcolor{Green}{Yes}
    4,     $id_5$,    products,    0.769,    {1, 3}          ,\textcolor{Green}{Yes}
    5,     $id_6$,    serving\_size,    0.742,    {3}          ,\textcolor{Green}{Yes}
    6,     $id_7$,    products,    0.739,    {1, 3}          ,\textcolor{Green}{Yes}
    7,     $id_8$,    products,    0.738,    {1, 3}          ,\textcolor{Green}{Yes}
    8,     $id_9$,    products,    0.737,    {1, 3}          ,\textcolor{Green}{Yes}
    9,     $id_{10}$,    serving\_size,    0.736,    {3}          ,\textcolor{Green}{Yes}
    10,     $id_{11}$,    nutrients,    0.733,              ,\textcolor{Red}{No}
    11,     $id_{12}$,    serving\_size,   0.732,    {3}          ,\textcolor{Green}{Yes}
    12,     $id_{13}$,    serving\_size,    0.732,    {3}          ,\textcolor{Green}{Yes}
    13,     $id_{14}$,    nutrients,    0.730,    {3}          ,\textcolor{Green}{Yes}
    14,     $id_{15}$,    serving\_size,    0.728,    {3}          ,\textcolor{Green}{Yes}
    15,     $id_{16}$,    products,    0.726,    {1, 3}          ,\textcolor{Green}{Yes}
    16,     $id_{17}$,    nutrients,    0.724,              ,\textcolor{Red}{No}
    17,     $id_{18}$,    nutrients,    0.723,              ,\textcolor{Red}{No}
    18,     $id_{19}$,    serving\_size,    0.722,    {3}          ,\textcolor{Green}{Yes}
    19,     $id_{20}$,    nutrients,    0.720,              ,\textcolor{Red}{No}

}
\end{adjustbox}
\textbf{\caption{\label{tab:experiment1-base-tables:analyst-tv}
                Experiment \ref{advanced-experiment:1} \textit{red gold} vs. related \textbf{base tables} with \textbf{Tuple Vectors} - an analyst's view.
                }
        }
\vspace{-7mm}
\end{table}\begin{table*}
\centering
\begin{adjustbox}{width=\textwidth,center}
\pgfplotstabletypeset[
    color cells={min=0.32,max=1.0}, format index, format stats-table2-exp3, format manufacturer, format method,
    col sep=comma,
    /pgfplots/colormap={orange}{
        rgb255(0cm)=(255,245,235);
        rgb255(1cm)=(253,141,60)},
    assign column name/.code=\pgfkeyssetvalue{/pgfplots/table/column
        name}{{\textbf{#1}}},
]{
    
    index,  Provenance Method,       manufacturer, name, Total, Length-L1, Length-L2, Length-L3,  {1.00$^{Top}$},            {0.75$^{Top}$},        {0.50$^{Top}$},        {0.25$^{Top}$}, Recall-L1, Recall-L2
    0, Tuple Vectors, conagra brands, healthy choice microwaveable chicken and rice soup, 1836, 2, 1834, 0, 0.397, 0.406, 0.417, 0.448, 0.5, 0.396
    1, , conagra brands, marie callenders cheesy chicken and rice dinners, 1836, 2, 1834, 0, 0.328, 0.345, 0.366, 0.433, 0.5, 0.328
    2, , general mills sales inc, annies homegrown organic vegan mac elbow rice pasta sauce, 2035, 2, 2033, 0, 0.408, 0.431, 0.435, 0.387, 0.5, 0.408
    3, , general mills sales inc, old el paso spanish rice, 2035, 2, 2033, 0, 0.331, 0.356, 0.393, 0.446, 0.5, 0.331
    0, Column Vectors, conagra brands, healthy choice microwaveable chicken and rice soup, 1836, 2, 1834, 0, 0.924, 0.931, 0.926, 0.924, 0.5, 0.924
    1, , conagra brands, marie callenders cheesy chicken and rice dinners, 1836, 2, 1834, 0, 0.925, 0.933, 0.935, 0.917, 0.5, 0.925
    2, , general mills sales inc, annies homegrown organic vegan mac elbow rice pasta sauce, 2035, 2, 2033, 0, 0.814, 0.818, 0.789, 0.750, 0.5, 0.814
    3, , general mills sales inc, old el paso spanish rice, 2035, 2, 2033, 0, 0.808, 0.806, 0.783, 0.743, 0.5, 0.808
}
\end{adjustbox}
\textbf{\caption{\label{tab:experiment3-materialized-views:precision-and-recall}
                Experiment \ref{advanced-experiment:3} query results vs. related materialized views ($\texttt{exp}\sb\texttt{2}$, \texttt{protein}, \texttt{sugars} and \texttt{unprepared}).
                }
        }
\vspace{-4mm}
\end{table*}
\end{experiment-withrun}

\begin{experiment-withrun}\label{advanced-experiment:3}
We ask for distinct pairs of (\texttt{manufacturer}, \texttt{name}) for products from the \texttt{sugars} materialized view, that contain \textit{rice} in their name, and are produced by manufacturers that appear in the $\texttt{exp}\sb\texttt{2}$ materialized view:

\begin{figure}[h]
\vspace{-2mm}
\raggedright
\begin{small}
    \texttt{
    \textcolor{white}{F}\sqlcolor{SELECT} p.\fieldcolor{manufacturer}, t.\fieldcolor{name}\\
    \textcolor{white}{FF}\sqlcolor{FROM} (\sqlcolor{SELECT DISTINCT} sgr.\fieldcolor{name} \\
    \textcolor{white}{FFFFFFFF}\sqlcolor{FROM} $\texttt{exp}\sb\texttt{2}$, sugars sgr\\
    \textcolor{white}{FFFFFFFF}\sqlcolor{WHERE} $\texttt{exp}\sb\texttt{2}$.\fieldcolor{manufacturer} = sgr.\fieldcolor{manufacturer}\\
    \textcolor{white}{FFFFFFFF}\sqlcolor{AND POSITION}(\strcolor{'rice'} \sqlcolor{IN} sgr.\fieldcolor{name}) > \numcolor{0}) t, products p\\
    \textcolor{white}{FF}\sqlcolor{WHERE} t.\fieldcolor{name} = p.\fieldcolor{name} \\ 
    \textcolor{white}{FF}\sqlcolor{GROUP BY} p.\fieldcolor{manufacturer}, t.\fieldcolor{name}
    }
\end{small}
\vspace{-2mm}
\end{figure}

\begin{table}
\centering
\begin{adjustbox}{width=\columnwidth,center}
\pgfplotstabletypeset[
    color cells={min=0.55,max=0.99, textcolor=white}, format index, format analyst, format Lineage,
    col sep=comma,
    /pgfplots/colormap={purple}{
        rgb255(0cm)=(239,237,245);
        rgb255(1cm)=(106,81,163)},
    assign column name/.code=\pgfkeyssetvalue{/pgfplots/table/column
        name}{{\textbf{#1}}},
]{
    index,     tuple-id, related-table,      {Similarity},      lineage-level,  Lineage 
   0,     $id_1$,    sugars,    0.981,    {1}          ,\textcolor{Green}{Yes}
    1,     $id_2$,    protein,     0.935,              ,\textcolor{Red}{No}
    2,     $id_3$,    $\texttt{exp}\sb\texttt{2}$,    0.935,    {1}          ,\textcolor{Green}{Yes}
    3,     $id_4$,    sugars,    0.934,              ,\textcolor{Red}{No}
    4,     $id_5$,    unprepared,    0.907,    {2}          ,\textcolor{Green}{Yes}
    5,     $id_6$,    protein,    0.906,             ,\textcolor{Red}{No}
    6,     $id_7$,    sugars,    0.891,             ,\textcolor{Red}{No}
    7,     $id_8$,    protein,    0.887,    {2}          ,\textcolor{Green}{Yes}
    8,     $id_9$,    sugars,    0.886,           ,\textcolor{Red}{No}
    9,     $id_{10}$,    unprepared,    0.886,      {2}        ,\textcolor{Green}{Yes}
    10,     $id_{11}$,    sugars,    0.886,              ,\textcolor{Red}{No}
    11,     $id_{12}$,    protein,   0.886,     {2}        ,\textcolor{Green}{Yes}
    12,     $id_{13}$,    protein,    0.884,    {2}          ,\textcolor{Green}{Yes}
    13,     $id_{14}$,    protein,    0.883,     {2}        ,\textcolor{Green}{Yes}
    14,     $id_{15}$,    protein,    0.883,        {2}      ,\textcolor{Green}{Yes}
}
\end{adjustbox}
\textbf{\caption{\label{tab:experiment3-materialized-views:analyst-cv}
                Experiment \ref{advanced-experiment:3} result tuple No.\texttt{2} vs. related \textbf{materialized views} with \textbf{Column Vectors} - an analyst's view.
                }
        }
\vspace{-4mm}
\end{table}
\begin{table}
\centering
\begin{adjustbox}{width=\columnwidth,center}
\pgfplotstabletypeset[
    color cells={min=0.55,max=0.98, textcolor=white}, format index, format analyst, format Lineage,
    col sep=comma,
    /pgfplots/colormap={purple}{
        rgb255(0cm)=(239,237,245);
        rgb255(1cm)=(106,81,163)},
    assign column name/.code=\pgfkeyssetvalue{/pgfplots/table/column
        name}{{\textbf{#1}}},
]{
    index,     tuple-id, related-table,      {Similarity},      lineage-level,  Lineage 
    0,     $id_1$,    sugars,    0.965,    {1}          ,\textcolor{Green}{Yes}
    1,     $id_2$,    protein,     0.964,              ,\textcolor{Red}{No}
    2,     $id_3$,    unprepared,    0.957,    {2}          ,\textcolor{Green}{Yes}
    3,     $id_4$,    sugars,    0.809,              ,\textcolor{Red}{No}
    4,     $id_5$,    $\texttt{exp}\sb\texttt{2}$,    0.809,    {1}          ,\textcolor{Green}{Yes}
    5,     $id_6$,    protein,    0.768,             ,\textcolor{Red}{No}
    6,     $id_7$,    unprepared,    0.755,         {2}    ,\textcolor{Green}{Yes}
    7,     $id_8$,    $\texttt{exp}\sb\texttt{2}$,    0.720,              ,\textcolor{Red}{No}
    8,     $id_9$,    sugars,    0.661,           ,\textcolor{Red}{No}
    9,     $id_{10}$,    sugars,    0.658,              ,\textcolor{Red}{No}
    10,     $id_{11}$,    sugars,    0.654,              ,\textcolor{Red}{No}
    11,     $id_{12}$,    protein,   0.653,     {2}        ,\textcolor{Green}{Yes}
    12,     $id_{13}$,    protein,    0.653,    {2}          ,\textcolor{Green}{Yes}
    13,     $id_{14}$,    protein,    0.653,     {2}        ,\textcolor{Green}{Yes}
    14,     $id_{15}$,    sugars,    0.653,              ,\textcolor{Red}{No}
}
\end{adjustbox}
\textbf{\caption{\label{tab:experiment3-materialized-views:analyst-tv}
                Experiment \ref{advanced-experiment:3} result tuple No.\texttt{2} vs. related \textbf{materialized views} with \textbf{Tuple Vectors} - an analyst's view.
                }
        }
\vspace{-7mm}
\end{table}

Table \ref{tab:experiment3-materialized-views:precision-and-recall} presents lineage related statistics, collected when computing the approximate lineage of four (out of nine, for brevity) of the result tuples in the output of this query against all the tuples from the related \textbf{materialized views}: $\texttt{exp}\sb\texttt{2}$, \texttt{protein}, \texttt{sugars} and \texttt{unprepared}, which contain $10,298$ tuples in total. Tables \ref{tab:experiment3-materialized-views:analyst-cv} and \ref{tab:experiment3-materialized-views:analyst-tv} present the top 15 tuples from the related materialized views, ranked by similarity to result tuple No.\texttt{2}, using the CV and TV methods, respectively. We make the following observations by analyzing the results:
\begin{enumerate*}
    \item Once again, the CV method demonstrates overall superiority, compared to the TV method, in terms of total precision and per-level recall.
    \item Analyzing result tuple No.\texttt{2}, the top rated tuple in both the CV and TV methods (see Tables \ref{tab:experiment3-materialized-views:analyst-cv} and \ref{tab:experiment3-materialized-views:analyst-tv}, respectively) is the only \texttt{sugars} tuple in the $1^{st}$ lineage level. The $2^{nd}$ tuple in the $1^{st}$ (L[1]) lineage level (from $\texttt{exp}\sb\texttt{2}$) is ``discovered'' earlier (ranked $3^{rd}$ vs. $5^{th}$) in the CV method, with no similarity score separation from the higher ranked tuples. By contrast, the TV method exhibits a significant similarity score drop all across the top 15 ranked tuples (0.97 for $1^{st}$ ranked vs. 0.81 for $5^{th}$ ranked vs. 0.65 for $15^{th}$ ranked). We view a more stable similarity score progression as indicating a better overall performance.
    \item Analyzing Tables \ref{tab:experiment3-materialized-views:precision-and-recall} and \ref{tab:experiment3-materialized-views:analyst-cv}, it seems the numbers for result No.\texttt{2} are not as good (compared to previous experiments and the result tuples No.\texttt{0} and No.\texttt{1} in this current experiment). A closer look reveals that out of the top 100 ranked tuples (by similarity) - all the mistakes are made on tuples from the \texttt{sugars} and \texttt{protein} materialized views. The \textcolor{Red}{No} tuples from \texttt{sugars} and \texttt{protein} are mostly products that appear also in the \texttt{unprepared} materialized view, and are actually in the lineage (i.e., \textcolor{Green}{Yes} tuples) of result tuple No.\texttt{2}. It seems that these \textcolor{Red}{No} tuples from \texttt{sugars} and \texttt{protein} have the same lineage vectors per the product related columns as the respective \textcolor{Green}{Yes} tuples from \texttt{unprepared}. Now, the Query Dependency DAG has no real filtering capabilities in this case (since the queries that created \texttt{sugars} and \texttt{protein} are relevant), hence we are not able to effectively eliminate these mistakes.

\end{enumerate*}
\end{experiment-withrun}

\begin{experiment-withrun}\label{advanced-experiment:2}
We ask for distinct pairs of (\texttt{manufacturer}, \texttt{name}) for products that are listed in the $\texttt{exp}\sb\texttt{4}$ materialized view:

\begin{figure}[h]
\raggedright
\vspace{-2mm}
\begin{small}
    \texttt{
    \textcolor{white}{F}\sqlcolor{SELECT} p.\fieldcolor{manufacturer}, exp$\sb\texttt{4}$.\fieldcolor{name}\\
    \textcolor{white}{FF}\sqlcolor{FROM} exp$\sb\texttt{4}$, products p\\
    \textcolor{white}{FF}\sqlcolor{WHERE} exp$\sb\texttt{4}$.\fieldcolor{name} = p.\fieldcolor{name} \\ 
    \textcolor{white}{FF}\sqlcolor{GROUP BY} p.\fieldcolor{manufacturer}, exp$\sb\texttt{4}$.\fieldcolor{name}
    }
\end{small}
\vspace{-2mm}
\end{figure}
\begin{table*}
\centering
\begin{adjustbox}{width=\textwidth,center}
\pgfplotstabletypeset[
    color cells={min=0.32,max=1.0}, format index, format stats-table1-exp2, format manufacturer, format method,
    col sep=comma,
    /pgfplots/colormap={orange}{
        rgb255(0cm)=(255,245,235);
        rgb255(1cm)=(253,141,60)},
    assign column name/.code=\pgfkeyssetvalue{/pgfplots/table/column
        name}{{\textbf{#1}}},
]{
    
    index,  Provenance Method,       manufacturer, name, Total, Length-L1, Length-L2, Length-L3, Length-L4,  {1.00$^{Top}$},            {0.75$^{Top}$},        {0.50$^{Top}$},        {0.25$^{Top}$}, Recall-L1, Recall-L3, Recall-L4
    0, Tuple Vectors, campbell soup company, v8 beverage carrot mango, 82, 2, 0, 4, 80, 0.195, 0.194, 0.238, 0.286, 1, 1, 0.175
    1, , campbell soup company, v8 vfusion beverage peach mango, 82, 2, 0, 4, 80, 0.183, 0.226, 0.262, 0.333, 1, 1, 0.163
    0, Column Vectors, campbell soup company, v8 beverage carrot mango, 82, 2, 0, 4, 80, 0.805, 0.774, 0.810, 0.905, 1, 0.5, 0.813
    1, , campbell soup company, v8 vfusion beverage peach mango, 82, 2, 0, 4, 80, 0.805, 0.774, 0.786, 0.810, 1, 0.5, 0.813
}
\end{adjustbox}
\textbf{\caption{\label{tab:experiment2-base-tables:precision-and-recall}
                Experiment \ref{advanced-experiment:2} query results vs. related base tables (\texttt{products}, \texttt{nutrients} and \texttt{serving\_size}).
                }
        }
\vspace{-4mm}
\end{table*}

Table \ref{tab:experiment2-base-tables:precision-and-recall} presents lineage related statistics, collected when computing the approximate lineage of the (two) result tuples in the output of this query ((\textit{campbell soup company, v8 beverage carrot mango}), (\textit{campbell soup company, v8 vfusion beverage peach mango})) against all the tuples from the related \textbf{base tables}: \texttt{products}, \texttt{nutrients} and \texttt{serving\_size}, which contain $29,322$ tuples in total. Tables \ref{tab:experiment2-base-tables:analyst-cv} and \ref{tab:experiment2-base-tables:analyst-tv} present a subset of the top 83 tuples from the related base tables, ranked by similarity to result tuple No.\texttt{0}, using the CV and TV methods, respectively. We make the following observations by analyzing the results:
\begin{enumerate*}
    \item The lineage sizes (of both result tuples) of the different levels (1-4) sum up to more than the total distant lineage sizes. E.g., looking at result tuple No.\texttt{0}: $2 + 0 + 4 + 80 > 82$. In this case, it means that at least two of the four tuples from the $3^{rd}$ lineage level (which contains tuples from the \texttt{products} and \texttt{serving\_size} relations) are also a part of either the $1^{st}$ lineage level (which contains tuples from the \texttt{products} relation), or the $4^{th}$ lineage level (which contains tuples from the \texttt{products}, \texttt{nutrients} and \texttt{serving\_size} relations).
    \item The Column Vectors (CV) method demonstrates overall superiority, compared to the Tuple Vectors (TV) method, topping at a precision score of 0.81 for the top 82 (total lineage size) approximate lineage tuples, for both result tuples.  Moreover, the TV method demonstrates unusually low  ($\approx0.2$) precision scores for this query.
    \item The TV method exhibits perfect L[1] and L[3] recall scores, and a low L[4] recall score, for both result tuples (see Table \ref{tab:experiment2-base-tables:precision-and-recall}). We conclude that the TV method is highly effective in finding the $1^{st}$ and $3^{rd}$ lineage-level tuples, but performs poorly in finding the $4^{th}$ lineage-level tuples, as is evidenced by the \textcolor{Red}{No} results in Table \ref{tab:experiment2-base-tables:analyst-tv}.
    \item The CV method exhibits relatively high L[1] and L[4] recall scores and a somewhat mediocre L[3] recall score for both result tuples (see Table \ref{tab:experiment2-base-tables:precision-and-recall}), evidenced by observing the Lineage Level(s) column in Table \ref{tab:experiment2-base-tables:analyst-cv}. That is, not only that almost all the top-20 tuples by similarity are really a part of the distant lineage, but also, the ranking preserves a level-based bias (the top-2 tuples are the only ones that appear both in the $1^{st}$, $3^{rd}$ and $4^{th}$ lineage levels).
    \item Analyzing the Tables \ref{tab:experiment2-base-tables:analyst-cv} and \ref{tab:experiment2-base-tables:analyst-tv}, we see that the two $3^{rd}$ lineage-level tuples from the \texttt{serving\_size} relation are discovered significantly earlier in the ranking by the TV method ($3^{rd}$ and $4^{th}$ tuples, by similarity) in comparison with the CV method ($80^{th}$ and $83^{rd}$ tuples, by similarity). This means that the TV method does a better job on the $3^{rd}$ lineage level, in terms of recall. This observation is backed up by the L[3] recall results we see in Table \ref{tab:experiment2-base-tables:precision-and-recall}.
    \item Notice the $4^{th}$ ranked tuple (by similarity) in Table \ref{tab:experiment2-base-tables:analyst-cv}, which is a \textcolor{Red}{No} lineage tuple, with a relatively high similarity score. A closer look reveals that this tuple is a product named \textit{v8 splash beverage mango peach}, which is also produced by \textit{campbell soup company}. Interestingly, it also appears as the first \textcolor{Red}{No} tuple when analyzing result tuple No.\texttt{1} vs. related base tables with the CV method (we do not show this explicitly).
    \item Note that a random choice of the top 82 lineage tuples would have yielded a $\frac{82}{29,322} \approx 0.003$ precision score, which is several orders of magnitude worse than the scores of the CV (and even the TV) methods. 
\end{enumerate*}

\begin{table}
\centering
\begin{adjustbox}{width=\columnwidth,center}
\pgfplotstabletypeset[
    color cells={min=0.7,max=0.9, textcolor=white}, format index, format analyst, format Lineage,
    col sep=comma,
    /pgfplots/colormap={purple}{
        rgb255(0cm)=(239,237,245);
        rgb255(1cm)=(106,81,163)},
    assign column name/.code=\pgfkeyssetvalue{/pgfplots/table/column
        name}{{\textbf{#1}}},
]{
    index,     tuple-id, related-table,      {Similarity},      lineage-level,  Lineage 
    0,     $id_1$,    products,    0.994,    {1, 3, 4}          ,\textcolor{Green}{Yes}
    1,     $id_2$,    products,     0.993,    {1, 3, 4}          ,\textcolor{Green}{Yes}
    2,     $id_3$,    products,    0.963,    {4}          ,\textcolor{Green}{Yes}
    3,     $id_4$,    products,    0.954,              ,\textcolor{Red}{No}
    4,     $id_5$,    products,    0.953,    {4}          ,\textcolor{Green}{Yes}
    5,     $id_6$,    products,    0.951,    {4}         ,\textcolor{Green}{Yes}
    $\vdots$,     ,    $\vdots$,    ,              $\vdots$,
    14,     $id_{15}$,    products,    0.943,    {4}          ,\textcolor{Green}{Yes}
    15,     $id_{16}$,    products,    0.942,              ,\textcolor{Red}{No}
    16,     $id_{17}$,    products,    0.942,    {4}          ,\textcolor{Green}{Yes}
    $\vdots$,     ,    $\vdots$,    ,              $\vdots$,
    79,     $id_{80}$,    serving\_size,    0.905,    {3}          ,\textcolor{Green}{Yes}
    $\vdots$,     ,    $\vdots$,    ,              $\vdots$,
    82,     $id_{83}$,    serving\_size,    0.893,    {3}          ,\textcolor{Green}{Yes}

}
\end{adjustbox}
\textbf{\caption{\label{tab:experiment2-base-tables:analyst-cv}
                Experiment \ref{advanced-experiment:2} result tuple No.\texttt{0} vs. related \textbf{base tables} with \textbf{Column Vectors} - an analyst's view.
                }
        }
\vspace{-7mm}
\end{table}
\begin{table}
\centering
\begin{adjustbox}{width=\columnwidth,center}
\pgfplotstabletypeset[
    color cells={min=0.7,max=0.8, textcolor=white}, format index, format analyst, format Lineage,
    col sep=comma,
    /pgfplots/colormap={purple}{
        rgb255(0cm)=(239,237,245);
        rgb255(1cm)=(106,81,163)},
    assign column name/.code=\pgfkeyssetvalue{/pgfplots/table/column
        name}{{\textbf{#1}}},
]{
    index,     tuple-id, related-table,      {Similarity},      lineage-level,  Lineage 
    0,     $id_1$,    products,    0.968,    {1, 3, 4}          ,\textcolor{Green}{Yes}
    1,     $id_2$,    products,     0.961,    {1, 3, 4}          ,\textcolor{Green}{Yes}
    2,     $id_3$,    serving\_size,    0.850,    {3}          ,\textcolor{Green}{Yes}
    3,     $id_4$,    serving\_size,    0.844,    {3}          ,\textcolor{Green}{Yes}
    4,     $id_5$,    nutrients,    0.826,    {4}          ,\textcolor{Green}{Yes}
    5,     $id_6$,    nutrients,    0.819,    {4}          ,\textcolor{Green}{Yes}
    6,     $id_7$,    nutrients,    0.818,              ,\textcolor{Red}{No}
    7,     $id_8$,    nutrients,    0.813,              ,\textcolor{Red}{No}
    8,     $id_9$,    nutrients,    0.810,              ,\textcolor{Red}{No}
    9,     $id_{10}$,    nutrients,    0.808,               ,\textcolor{Red}{No}
            $\vdots$,     ,    $\vdots$,    ,              $\vdots$,
    18,     $id_{19}$,    nutrients,    0.798,              ,\textcolor{Red}{No}
    19,     $id_{20}$,    nutrients,    0.797,              ,\textcolor{Red}{No}

}
\end{adjustbox}
\textbf{\caption{\label{tab:experiment2-base-tables:analyst-tv}
                Experiment \ref{advanced-experiment:2} result tuple No.\texttt{0} vs. related \textbf{base tables} with \textbf{Tuple Vectors} - an analyst's view.
                }
        }
\vspace{-7mm}
\end{table}

\end{experiment-withrun}

\section{Conclusions}
\label{chap:conclusions}

We present a novel approach, in which we approximate lineage (a specific type of provenance) using embedding techniques; drawing inspiration from word vectors in the NLP domain. Our methods consume a reasonable constant amount of space per DB tuple. Finally, we showed experimental results of our two approximate lineage computation methods, TV and CV, and provided insights as to their performance when compared to the exact lineage obtained from the  ProvSQL system. 
The examples we presented suggest a high usefulness potential for the proposed approximate lineage methods and the further suggested enhancements.
This especially holds for the Column Vectors method which exhibits high precision and per-level recall.
\par Our methods implicitly produce a ``natural ranking'' of explanations. It is unclear how to get this kind of ranking from semiring provenance polynomials (it might require significant additional work). There is a work by Deutch et al. \cite{Deutch2015} that ranks derivation trees for Datalog programs; the ranking is  based on weights assigned to the deriving Datalog rules and tuples from the underlying database. In contrast, we deal with simpler SQL queries and producing ranking does not require additional work.

Comparisons to other, possibly simpler methods, is needed; yet, obvious ideas seem inferior to word embedding. For example, one could think of achieving the same functionality using simple data-type dependent featurization such as numerical scaling and one-hot encoding.
This ``simpler approach'' makes assumptions about the problem domain, e.g., the vocabulary. Thus, rendering the update of such vocabulary impractical, as opposed to word embeddings via \textsc{Word2Vec} that can be modified to support incremental training \cite{w2v_incremental}.
Most importantly, word embeddings support similarity queries by grouping closely ``semantically'' similar words in a low dimensional space. This property lies at the heart of our lineage-approximation work, as it enables encoding lineage in a low-dimensional space.

\bibliographystyle{ACM-Reference-Format}
\bibliography{back/general}

\end{document}